\documentclass[11pt,a4paper]{article}
\pdfoutput=1

\usepackage[utf8]{inputenc}
\usepackage{a4wide}
\usepackage{amsmath}
\usepackage{cite}
\usepackage{xcolor}
\usepackage{amssymb}
\usepackage{amsfonts}
\usepackage{booktabs}
\usepackage{makecell}
\usepackage{multirow}
\usepackage{pdflscape}
\usepackage{textcomp}
\usepackage{gensymb}
\usepackage{bigstrut}
\usepackage{array}
\usepackage{enumerate}
\usepackage[small,bf]{caption}
\usepackage{subcaption}
\usepackage{diagbox}
\usepackage{graphicx}
\usepackage{hyperref}
\usepackage{mathbbol}
\usepackage{appendix}
\usepackage{verbatim}
\usepackage{geometry}
\usepackage{makecell}
\usepackage{listings}
\usepackage{mathtools}
\usepackage{dsfont}
\usepackage{accents}
\usepackage{longtable}
\usepackage{threeparttable}
\usepackage{floatpag}

\usepackage[capitalise]{cleveref} 

\makeatletter
\newcommand {\crefext}[2]{\csname cref@#1@format\endcsname{#2}{}{}}
\newcommand {\Crefext}[2]{\csname Cref@#1@format\endcsname{#2}{}{}}
\makeatother

\DeclareMathOperator{\diag}{diag}

\DeclareMathOperator{\im}{Im}
\DeclareMathOperator{\re}{Re}

\def\aprop{
  \def\p{%
    \setbox0=\vbox{\hbox{$\propto$}}%
    \ht0=0.6ex \box0 }%
  \def\s{%
    \vbox{\hbox{$\sim$}}%
  }%
  \mathrel{\raisebox{0.7ex}{%
      \mbox{$\underset{\s}{\p}$}%
    }}%
}
\def\ato{\,  \xrightarrow[ST\text{-diag.}]{\,\,\aprop\,\,}\,}

\newcommand{\nine}{\textcolor{red}{9+}}
\newcommand{\five}{\textcolor{orange}{5+}}

\newcommand{\id}{\mathds{1}}
\newcommand{\ta}{\tilde\alpha}

\newcommand*{\belowrulesepcolor}[1]{%
  \noalign{%
    \kern-\belowrulesep
    \begingroup
      \color{#1}%
      \hrule height\belowrulesep
    \endgroup
  }%
}
\newcommand*{\aboverulesepcolor}[1]{%
  \noalign{%
    \begingroup
      \color{#1}%
      \hrule height\aboverulesep
    \endgroup
    \kern-\aboverulesep
  }%
}

\allowdisplaybreaks
\numberwithin{equation}{section}

\begin{document}

%%%%%%%%%%%%%%%%%%%%%%%
\begin{titlepage}

\vspace*{-15mm}
\begin{flushright}
SISSA 11/2023/FISI\\
CFTP/23-002
\end{flushright}
\vspace*{5mm}

\begin{center}
{\bf\LARGE {Quarks at the modular $S_4$ cusp}
}\\[8mm]
I.~de~Medeiros~Varzielas\(^{\,a,}\)\footnote{E-mail: \texttt{ivo.de@udo.edu}},
M.~Levy\(^{\,a,}\)\footnote{E-mail: \texttt{miguelplevy@ist.utl.pt}},
J.~T.~Penedo\(^{\,a,}\)\footnote{E-mail: \texttt{joao.t.n.penedo@tecnico.ulisboa.pt}},
S.~T.~Petcov\(^{\,b,c,}\)\footnote{Also at
Institute of Nuclear Research and Nuclear Energy,
Bulgarian Academy of Sciences, 1784 Sofia, Bulgaria.}\\
 \vspace{5mm}
\(^{a}\)\,{\it CFTP, Departamento de Física, Instituto Superior Técnico, Universidade de Lisboa,\\
Avenida Rovisco Pais 1, 1049-001 Lisboa, Portugal} \\
\vspace{2mm}
\(^{b}\)\,{\it SISSA/INFN, Via Bonomea 265, 34136 Trieste, Italy} \\
\vspace{2mm}
\(^{c}\)\,{\it Kavli IPMU (WPI), UTIAS, The University of Tokyo, \\
Kashiwa, Chiba 277-8583, Japan}
\end{center}
\vspace{2mm}

\begin{abstract}

We analyse the possibility of describing quark masses, mixing and CP violation in $S'_4$ modular flavour models without flavons. We focus on the case where the closeness of the modulus to the point of residual $\mathbb{Z}^{ST}_3$ symmetry (the cusp) plays a role in generating quark mass hierarchies and discuss the role modular form normalisations play in such constructions. We find that fitting quark data requires explicit CP breaking, unless a second modulus is introduced. 

\end{abstract}

\end{titlepage}
\setcounter{footnote}{0}
%%%%%%%%%%%%%%%%%%%%%%%
%

%%%%%%%%%%%%%%%%%%%%%%%%%%%%%%%%%%%%%%%%%%%%%%
\section{Introduction}
\label{sec:intro}
%%%%%%%%%%%%%%%%%%%%%%%%%%%%%%%%%%%%%%%%%%%%%%

In an era of increasing experimental precision, the flavour puzzle stands as an enigma that hints at new physics beyond the Standard Model (SM). The observed hierarchies among the masses of the three generations of up quarks, down quarks and charged leptons suggest the action of a mechanism which is not yet understood. The peculiar and seemingly unrelated mixing patterns in the quark and lepton sectors add to the mystery. 
Overall, one counts 22 low-energy independent parameters contained in the Yukawa sector of the SM extended with massive Majorana neutrinos. These are 12 masses, 6 mixing angles and 4 CP-violating (CPV) phases. The quest for a principle, akin to the gauge principle, which economically describes this plethora of parameters is enticing and has led to the development of the discrete symmetry approach to flavour~\cite{Ishimori:2010au}, largely focused on the leptonic sector~\cite{Altarelli:2010gt,King:2014nza,Tanimoto:2015nfa,Petcov:2017ggy,Feruglio:2019ybq}.

In the past years, a new avenue in model building has been opened by the proposal of using modular invariance as a flavour symmetry~\cite{Feruglio:2017spp} (see~\cite{Kobayashi:2023zzc} for a recent review).%
\footnote{Modular invariance may also play a role in solving the strong CP problem~\cite{Feruglio:2023uof}.}
In this supersymmetric (SUSY) framework, the components of Yukawa and mass matrices may be obtained from modular forms of level \(N\) as well as from a small set of coupling constants in the superpotential. It is the holomorphicity of the latter which allows for a predictive setup. In the simplest case, these forms are functions of a single complex scalar field, the modulus \(\tau\).
The theory is assumed to be invariant under the whole modular group \(\Gamma \equiv SL(2,\mathbb{Z})\). Matter fields, however, transform in representations of a finite inhomogeneous (homogeneous) modular group \(\Gamma^{(\prime)}_N\), which plays the role of a flavour symmetry. The finite groups \(\Gamma_N\) are isomorphic to the permutation groups \( S_3\), \( A_4\), \( S_4\) and \( A_5\)~\cite{deAdelhartToorop:2011re} typically used in model building, while the groups \(\Gamma_N'\) are isomorphic to the corresponding ``double covers''. 
 
No flavons are required in modular flavour models, since the vacuum expectation value (VEV) of \(\tau\) may be the only source of symmetry breaking, fixing the values of the modular forms and, consequently, the flavour structure of fermion mass matrices.
Moreover, the VEV of \(\tau\) may also be the only source of breaking of a generalised CP (gCP) symmetry, which can be consistently combined with the modular symmetry~\cite{Novichkov:2019sqv} (see also~\cite{Baur:2019kwi}).
Note that any VEV for \(\tau\) breaks the full modular symmetry. A remnant symmetry may yet be preserved, but only at one of three fixed points, \(\tau_\text{sym} = i,\, \omega, \,i \infty\)~\cite{Novichkov:2018ovf}, with \(\omega \equiv \exp(2\pi i / 3)\) --- the (left) cusp. For each of these cases, a residual \(\mathbb{Z}^{S}_2\), \(\mathbb{Z}^{ST}_3\), or \(\mathbb{Z}^{T}_N\) symmetry, respectively, is left unbroken.%
\footnote{In the case of the finite groups \(\Gamma'_N\), an extra residual \(\mathbb{Z}^R_2\) is preserved~\cite{Novichkov:2020eep}.}

Another unique characteristic of the modular framework is the fact that fermion mass hierarchies may follow from the properties of the modular forms, without fine-tuning~\cite{Novichkov:2021evw}.
This mechanism was recently employed in Refs.~\cite{Petcov:2022fjf,Kikuchi:2023cap,Abe:2023ilq,Kikuchi:2023jap,Abe:2023qmr,Petcov:2023vws,Abe:2023dvr} and will also be explored in the present work. It relies on the VEV of the modulus taking a value close to one of the symmetric points.
Indeed, fermion mass matrices are strongly constrained at or in the vicinity of the points of residual symmetry (see also~\cite{Novichkov:2018ovf,Novichkov:2018nkm,Novichkov:2018yse,Okada:2020ukr,Feruglio:2021dte,Kikuchi:2022svo,Hoshiya:2022qvr,Feruglio:2023bav,Feruglio:2023mii}).
As an example, for the viable fine-tuning-free model presented in Ref.~\cite{Novichkov:2021evw}, \(\tau\) is driven by lepton data to the vicinity of the cusp \(\omega\). The numerical fit region corresponds to a ring around the latter, with a small radius \(|u| \simeq 0.007\), where \(u \equiv (\tau-\omega)/(\tau-\omega^2)\). While the best-fit point \(\tau \simeq -\,0.496 + 0.877\,i\) follows from a fit of low-energy lepton data, such peculiar values of \(\tau\) may instead be selected, in a top-down approach, by a dynamical principle~\cite{Novichkov:2022wvg,Knapp-Perez:2023nty}.%
\footnote{
For explanations of the fermion mass hierarchies relying on extra (weighted) scalars, with modular weights playing the role of Froggatt-Nielsen charges\mbox{\cite{Froggatt:1978nt}}, see instead\mbox{\cite{Criado:2019tzk,King:2020qaj,Kuranaga:2021ujd}}.
}

At present, a vast number of modular flavour models can be found in the literature, including:
\begin{itemize}
    \item models of flavour based on the groups 
\(\Gamma_2 \simeq S_3\)~\cite{Kobayashi:2018vbk,Okada:2019xqk,Mishra:2020gxg,Meloni:2023aru},
\(\Gamma_3 \simeq A_4\)~\cite{Criado:2018thu,Kobayashi:2018scp, Novichkov:2018yse, Okada:2018yrn, Nomura:2019yft, Okada:2019mjf, Nomura:2019jxj, Ding:2019zxk, Kobayashi:2019gtp,Nomura:2019lnr,Asaka:2019vev,Ding:2019gof,Zhang:2019ngf,Nomura:2019xsb,Wang:2019xbo,Hutauruk:2020xtk,Okada:2020dmb,Ding:2020yen,Behera:2020sfe,Nomura:2020cog,Behera:2020lpd,Asaka:2020tmo,Nagao:2020snm,Abbas:2020vuy,Kashav:2021zir,Okada:2021qdf,Tanimoto:2021ehw,Nagao:2021rio,Kobayashi:2021ajl,Dasgupta:2021ggp,Nomura:2021pld,Kobayashi:2021jqu,Okada:2021aoi,Otsuka:2022rak,Ahn:2022ufs,Nomura:2022hxs,Kobayashi:2022jvy,Kobayashi:2021pav,Kang:2022psa,Nomura:2022boj,Kim:2023jto,Devi:2023vpe,Mishra:2023ekx},
\(\Gamma_4 \simeq S_4\)~\cite{Penedo:2018nmg,Novichkov:2018ovf, Criado:2019tzk, Kobayashi:2019mna,Kobayashi:2019xvz,Okada:2019lzv,Ding:2019gof,Wang:2019ovr,Wang:2020dbp,Gehrlein:2020jnr,Nomura:2021ewm},
\(\Gamma_5 \simeq A_5\)~\cite{Novichkov:2018nkm, Ding:2019xna, Criado:2019tzk, Gehrlein:2020jnr},
and \(\Gamma_7 \simeq PSL(2,\mathbb{Z}_7)\)~\cite{Ding:2020msi},
\item models of flavour based on the ``double cover'' groups \(\Gamma'_3 \simeq T'\)~\cite{Liu:2019khw,Okada:2022kee,Ding:2022aoe,Mishra:2023cjc,Ding:2023ynd}, \(\Gamma'_4 \simeq S'_4\)~\cite{Novichkov:2020eep,Liu:2020akv,Ding:2022nzn} and \(\Gamma'_5 \simeq A'_5\)~\cite{Wang:2020lxk,Yao:2020zml,Behera:2021eut,Behera:2022wco} and on \(\Gamma'_6 \simeq S_3 \times T'\)~\cite{Li:2021buv,Abe:2023dvr} (see also~\cite{Liu:2021gwa}, considering a generalisation to other finite modular groups),
\item models exploring the interplay of modular and gCP symmetries~\cite{Kobayashi:2019uyt,Okada:2020brs,Yao:2020qyy,Wang:2021mkw,Ding:2021iqp,Qu:2021jdy},
\item models of quark-lepton 
unification~\cite{Kobayashi:2018wkl,Okada:2019uoy,Kobayashi:2019rzp,Lu:2019vgm,Abbas:2020qzc,Okada:2020rjb,Liu:2020akv,Yao:2020zml,Du:2020ylx,Yao:2020qyy,Zhao:2021jxg,Chen:2021zty,King:2021fhl,Ding:2021zbg,Qu:2021jdy,Nomura:2021yjb,Ding:2021eva,Nomura:2022boj,Ding:2022bzs,Du:2022lij,Benes:2022bbg,Nomura:2023kwz,Abe:2023qmr,Abe:2023dvr},
\item models with multiple moduli,
first considered phenomenologically in~\cite{Novichkov:2018ovf,Novichkov:2018yse} and further developed in~\cite{deMedeirosVarzielas:2019cyj,King:2019vhv,deMedeirosVarzielas:2020kji,Ding:2020zxw,deMedeirosVarzielas:2021pug,Kobayashi:2021pav,deMedeirosVarzielas:2022ihu,deMedeirosVarzielas:2022fbw,Abbas:2022slb,deAnda:2023udh}, and
\item models relating modular flavour symmetries and inflation~\cite{Gunji:2022xig,Abe:2023ylh}.
\end{itemize}
The outcomes of this (mostly) bottom-up strategy should eventually be linked with the top-down results of UV-complete theories~\cite{Kobayashi:2018rad, Kobayashi:2018bff, deAnda:2018ecu, Baur:2019kwi, Kariyazono:2019ehj, Baur:2019iai, Nilles:2020nnc, Kobayashi:2020hoc, Abe:2020vmv, Ohki:2020bpo, Kobayashi:2020uaj, Nilles:2020kgo, Kikuchi:2020frp, Nilles:2020tdp, Kikuchi:2020nxn, Baur:2020jwc, Ishiguro:2020nuf, Nilles:2020gvu, Ishiguro:2020tmo, Hoshiya:2020hki, Baur:2020yjl, Kikuchi:2021ogn, Almumin:2021fbk,Tatsuta:2021deu,Baur:2021mtl,Nilles:2021glx,Ishiguro:2021ccl,Baur:2021bly,Novichkov:2022wvg,Kikuchi:2022txy,Ishiguro:2022pde,Kikuchi:2022geu,Baur:2022hma,Kikuchi:2023clx,Knapp-Perez:2023nty,Kikuchi:2023awm,Kai:2023ivp,Kikuchi:2023uqo}, building towards a more predictive setup (see also~\cite{Chen:2019ewa,Chen:2021prl}).

Many of the models built so far have focused on the lepton sector, while fitting the quark sector --- either independently of leptons or in an unified manner --- has proven to be a challenge (see, e.g.,~\cite{Petcov:2023vws}). 
If no explanation for hierarchical parameters is sought, modular models have been found to fit the 10 quark sector parameters (6 masses, 3 angles, 1 CPV phase) using a minimum of 9 real parameters. This has been achieved, using fine-tuned parameters, for the modular groups \(S_4'\)~\cite{Liu:2020akv}, \(A_4\)~\cite{Yao:2020qyy} (with a marginal fit), and \(S_4\)~\cite{Qu:2021jdy}.
Instead, as mentioned above, there is a recent effort to additionally derive the quark mass hierarchies from the closeness of \(\tau\) to a symmetric point, via the mechanism of Ref.~\cite{Novichkov:2021evw}.
In this context, which allows one to avoid fine-tuning, one has so far considered:
\begin{itemize}
\item the modular group \(A_4\), with \(\tau \simeq \omega\)~\cite{Petcov:2022fjf} or \(\tau \simeq i\infty\)~\cite{Petcov:2023vws},
\item the modular group \(S_4'\), with \(\tau \simeq i \infty\)~\cite{Abe:2023ilq,Abe:2023qmr},
\item the modular group \(\Gamma_6 \simeq S_3 \times A_4\), with \(\tau \simeq i \infty\)~\cite{Kikuchi:2023cap}, 
\item the modular group \(\Gamma_6' \simeq S_3 \times A_4'\), with \(\tau \simeq i \infty\)~\cite{Abe:2023dvr}, and
\item the multiple modular group \(A_4\times A_4\times A_4\), assuming a common VEV \(\tau\) for the three moduli, with either \(\tau \simeq \omega\) or \(\tau \simeq i \infty\)~\cite{Kikuchi:2023jap}.%
\footnote{Refs.~\cite{Abe:2023qmr} and~\cite{Abe:2023dvr} also explore quark-lepton unification, from a bottom-up and grand unified theory perspective, respectively.}
\end{itemize}
With the exception of Ref.~\cite{Kikuchi:2023cap}, which relies on discrete choices for the parameters, these models can fit the quark data with a minimum of 11 real parameters (case \(p=1\) in Ref.~\cite{Abe:2023ilq}), in a {\it phenomenological} approach. Here and in what follows, ``phenomenological'' refers to the fact that some parameters are chosen to be real, despite the fact that they are complex in general --- their reality is not guaranteed by e.g.~a {\it consistent} combination of modular and gCP symmetries. Therefore, taking into account the ignored (but allowed) phases, the number of real parameters rises to a minimum of 13 in these models. 
It is hoped that a modular model can be built with a number of parameters coming close to the previously-found minimum of 9, while explaining the quark mass hierarchies.

\vskip 2mm
In this work, we look into the restrictions on viable and minimal \(S_4^{(\prime)}\) modular flavour models of the quark sector where the proximity of the modulus to the point of residual \(\mathbb{Z}_3^{ST}\) symmetry (the cusp) plays a fundamental role in determining quark mass hierarchies.
Although we focus on the quark case, all analytical results are also applicable to other fermions, e.g.~the charged leptons.
In~\cref{sec:framework} we expose our rationale, identifying the mass matrices which i) are minimal in terms of parameters, and ii) are hierarchical in a way which may be attributed to the properties of modular multiplets in the vicinity of the cusp.
These matrices are expanded and explored analytically in~\cref{sec:matrices}. Our numerical results, showing which models can be fitted to data, are collected in~\cref{sec:results}, while some illustrative benchmarks are discussed in~\cref{sec:natural}. We summarise and conclude in~\cref{sec:summary}.

%%%%%%%%%%%%%%%%%%%%%%%%%%%%%%%%%%%%%%%%%%%%%%
\section{Framework}
\label{sec:framework}
%%%%%%%%%%%%%%%%%%%%%%%%%%%%%%%%%%%%%%%%%%%%%%

%%%%%%%%%%%%%%%%%%%%%%%
\subsection{Modular symmetry as a flavour symmetry}
\label{sec:modframework}
%%%%%%%%%%%%%%%%%%%%%%%

For a given element \(\gamma\) of the modular group \(\Gamma\), with generators
\begin{equation}
  \label{eq:STR_def}
  S =
  \begin{pmatrix}
    0 & 1 \\ -1 & 0
  \end{pmatrix}
  \,, \quad
  T =
  \begin{pmatrix}
    1 & 1 \\ 0 & 1
  \end{pmatrix}
  \,, \quad
  R =
  \begin{pmatrix}
    -1 & 0 \\ 0 & -1
  \end{pmatrix}\,,
\end{equation}
obeying \(S^2 = R\), \((ST)^3 = R^2 = \id\), and \(RT = TR\), the modulus \(\tau\) transforms via fractional linear transformations, as
\begin{equation}
  \label{eq:tau_mod_trans}
  \gamma =
  \begin{pmatrix}
    a & b \\ c & d
  \end{pmatrix}
  \in \Gamma : \quad
  \tau \to \gamma \tau = \frac{a\tau + b}{c\tau + d} \,,
\end{equation}
while matter superfields \(\psi_i\) transform as weighted multiplets~\cite{Ferrara:1989bc,Ferrara:1989qb,Feruglio:2017spp},
\begin{equation}
  \label{eq:psi_mod_trans0}
  \psi_i \to (c\tau + d)^{-k} \, \rho_{ij}(\gamma) \, \psi_j \,,
\end{equation}
where \(\rho\) is a representation of~\(\Gamma\) and \(k\) is the modular weight of \(\psi\).
To employ the modular symmetry as a flavour symmetry, we start by fixing an integer level \(N > 1\) and assuming that \(\rho(\gamma) = \id\) for elements of the principal congruence subgroup \(\Gamma(N)\).
Hence, \(\rho\) is an ``almost trivial'' representation of the full modular group and a unitary representation of the finite quotient \(\Gamma_N' \equiv \Gamma \, \big/ \, \Gamma(N) \simeq SL(2, \mathbb{Z}_N)\). 
Moreover, if matter fields transform trivially under \(R\),
it is effectively a representation of the smaller group \(\Gamma_N \equiv \Gamma \, \big/ \left\langle \, \Gamma(N) \cup \mathbb{Z}_2^R \, \right\rangle\).

By requiring the invariance of the superpotential under modular transformations,
one finds that couplings \(Y_{I_1 \ldots I_n}(\tau)\) appearing in terms of the type \(\psi_{I_1}\ldots\psi_{I_n}\) must be special holomorphic functions of~\(\tau\) --- modular forms of level \(N\) --- obeying
\begin{equation}
  \label{eq:Y_mod_trans}
  Y_{I_1 \ldots I_n}(\tau) \,\xrightarrow{\gamma}\, Y_{I_1 \ldots I_n}(\gamma \tau) = (c\tau + d)^{k_Y} \rho_Y(\gamma) \,Y_{I_1 \ldots I_n}(\tau) \,.
\end{equation}
Modular forms carry weights \(k_Y = k_{I_1} + \ldots + k_{I_n}\) and furnish unitary representations \(\rho_Y\) of the finite modular group such that \(\rho_Y \,\otimes\, \rho_{I_1} \,\otimes \ldots \otimes\, \rho_{I_n} \supset \mathbf{1}\).
Modular symmetry may thus constrain the Yukawa structures of a model in a predictive way, since
the modular forms span finite-dimensional linear spaces of relatively low dimensionalities, for small values of~\(k\) and \(N\). 
We will focus on the case \(N=4\) and \(\Gamma'_4 \simeq S'_4\), although our reasoning can be straightforwardly extended to the other finite modular groups.
The group theory and modular forms for \(S_4'\) are summarised in~\cref{app:modS4p}.
Crucially, these forms can be written in terms of only two functions,
\(\theta\) and \(\varepsilon\)~\cite{Novichkov:2020eep}, defined by
\begin{equation}
  \begin{aligned}
    \theta(\tau) \,&\equiv\, 1
    + 2\sum_{k=1}^{\infty} q^{(2k)^2}
    \,=\, 1 + 2 q^4 + 2 q^{16} + \ldots \,, \\
    \varepsilon(\tau) \,&\equiv\, 2\sum_{k=1}^{\infty} q^{(2k-1)^2} \,=\, 2 q + 2 q^9 + 2 q^{25} + \ldots \,,
  \end{aligned}
  \label{eq:theta_eps_qexp}
\end{equation}
with \(q \equiv \exp(\pi i \tau / 2)\), and satisfying \(\varepsilon(\omega) / \theta(\omega) = (1-i)/(1+\sqrt{3})\).

\vskip 2mm
The VEV of \(\tau\) is restricted to the upper half-plane and plays the role of a spurion, parameterising all modular symmetry breaking in the absence of flavons. 
In such a case, the value of \(\tau\) can always be restricted to the fundamental domain \(\mathcal{D}\) of the modular group \(\Gamma\)~\cite{Novichkov:2018ovf}.
In a CP-invariant modular theory, an additional \(\mathbb{Z}^\text{CP}_2\) symmetry is preserved for \(\re\tau =0\) or for \(\tau\) on the border of \(\mathcal{D}\), while is broken at generic values of \(\tau\)~\cite{Novichkov:2019sqv,Novichkov:2020eep}. All three symmetric points  \(\tau_\text{sym} = i,\, \omega, \,i \infty\) preserve the CP symmetry. Finally, a \(\mathbb{Z}_2^R\) symmetry is always preserved, as the \(R\) generator is unbroken for any value of \(\tau\).

\vskip 2mm
At each of the symmetric points \(\tau_\text{sym} = i,\, \omega, \,i \infty\), flavour textures can be further constrained by the residual symmetry group, which may enforce the presence of multiple zero entries in the mass matrices.
As \(\tau\) moves away from the symmetric value, these entries will generically become non-zero.
The magnitudes of such (residual-)symmetry-breaking entries are controlled by the size of the departure \(\epsilon\) of \(\tau\) from \(\tau_\text{sym}\) and
by the field transformation properties under the residual symmetry group, which may depend on modular weights~\cite{Novichkov:2021evw}. 
Indeed, the zero entries of fermion mass matrices are expected to become \(\mathcal{O}(|\epsilon|^l)\).
The exponents \(l\) are extracted from products of factors which correspond to representations of the residual symmetry group 
and thus are not additional independent parameters (see Ref.~\cite{Novichkov:2021evw} for further details).

For the left cusp and \(N=4\), one may choose the small-magnitude parameter \(\epsilon\) as
\begin{equation}
\epsilon(\tau) \equiv 1- \frac{1+\sqrt{3}}{1-i}\frac{\varepsilon(\tau)}{\theta(\tau)}\,,
\quad\text{such that }\epsilon(\omega)=0\,.
\label{eq:epsilon}
\end{equation}
This parameter can be related to the previously-defined \(u = (\tau-\omega)/(\tau-\omega^2)\), via a \(u\)-expansion~\cite{Novichkov:2022wvg}, resulting in \(\epsilon \simeq 2.82\, u\) --- an approximation valid in the vicinity of the left cusp. 
Both \(|\epsilon|\) and \(|u|\) thus quantify the deviation of \(\tau\) from the point of residual \(\mathbb{Z}_3^{ST}\) symmetry.
Note that the right cusp, \(\omega + 1 = \exp(\pi i/3)\), is equivalent to the left cusp since they are related by the modular \(T\) transformation.
Additionally, the vicinity of the right cusp can be mapped to the vicinity of the left cusp \(\omega\), by the inverse of the \(T\) transformation, without affecting observables. While the resulting points may lie outside of the fundamental domain \(\mathcal{D}\), the corresponding \(|u|\) will be small and the results of Ref.~\cite{Novichkov:2021evw} apply.
Finally, note that if a fit of masses and mixing is possible in the vicinity of the right cusp within \(\mathcal{D}\), then such a fit is also possible in the vicinity of the left cusp within \(\mathcal{D}\), with \(\tau \to -\tau^*\), using the conjugated values of the superpotential constants, leading only to a sign flip of CPV phases~\cite{Novichkov:2018ovf}.

%%%%%%%%%%%%%%%%%%%%%%%
\subsection{Quark mass matrices}
\label{sec:matrix_th}
%%%%%%%%%%%%%%%%%%%%%%%

We are interested in finding quark mass matrices \(M\) whose singular values --- the quark masses --- are hierarchical as a consequence of the proximity of \(\tau\) to the cusp \(\omega\), i.e.~due to the smallness of \(\epsilon\) (the absolute value is implied, unless stated otherwise). As shown in~Ref.~\cite{Novichkov:2021evw}, for \(\tau \simeq \omega\) the only possible hierarchical massive spectra are of the type \(1:\epsilon:\epsilon^2\), since no higher powers of \(\epsilon\) are attainable.
The irrep pairs leading to such spectra have been identified therein for the case of a common weight across generations of a given isospin multiplet. Here, we lift this requirement and allow for different weights across generations.

Quark masses enter the superpotential \(W\) via a modular-invariant bilinear,
\begin{equation}
    W \,\supset\, 
    Q_i \, M(\tau)_{ij} \, q^c_j\,,
\end{equation}
where \(i,j=1,\ldots,3\) and we adopt a left-right convention. The quark superfield doublets \(Q \sim \bigoplus_\alpha (\mathbf{r}_\alpha, k_{\alpha})\) and quark superfield singlets \(q^c \sim \bigoplus_\beta (\mathbf{r}^c_\beta, k^c_{\beta})\) of a given sector (up or down) transform according to
\begin{equation}
\begin{aligned}
Q_\alpha \,&\xrightarrow{\gamma} \, 
(c\tau + d)^{-k_\alpha}\, \rho_{\mathbf{r}_\alpha}(\gamma)\, Q_\alpha\,,\\
q^c_\beta \,&\xrightarrow{\gamma} \, 
(c\tau + d)^{-k_\beta^c}\, \rho_{\mathbf{r}^c_\beta}(\gamma)\, q^c_\beta\,,
\end{aligned}
\end{equation}
under a modular transformation \(\gamma \in \Gamma\),
with \(\alpha\) and \(\beta\) labelling different irreps and weights.
To each pair \((\alpha,\beta)\) corresponds a sub-block of the matrix \(M\). A finite number of modular forms of weight \(k_\alpha+k^c_\beta\) may contribute to it.

\subsubsection{The issue of normalisations}
\label{sec:norms}
Let us comment on the relative size of modular forms contributing to \(M\). Taking an agnostic point of view, {\it we do not attribute physical significance to the absolute normalisation of modular form multiplets} in the vicinity of the cusp.
Indeed, note that modular forms always appear together with superpotential constants that can absorb their normalisation. Therefore, one can only meaningfully discuss the magnitude of these constants after some normalisation has been fixed for the modular forms.
In practice, in this work we will choose to normalise modular multiplets  at the fit value of \(\tau\) using the Euclidean norm. 
From the bottom-up perspective, constrained by group theory alone, this is a valid (albeit arbitrary) choice of normalisation.

To guarantee that fermion mass textures and hierarchies in our setup originate from the properties of modular forms,
they should then originate from the relations between the entries of a modular form multiplet --- the direction of the ``vector'' --- and not from the relative norms of independent modular forms --- the (arbitrary) sizes of the ``vectors''. So, a priori, different sub-blocks of \(M\) may contribute on the same footing to the mass matrix, and should not be \(\epsilon\)-suppressed among themselves.%
\footnote{This philosophy differs from what is considered in Refs.~\cite{Kikuchi:2023cap,Abe:2023ilq,Kikuchi:2023jap}.}

To illustrate this last point, consider as an example \(Q \sim (\mathbf{2},0) \oplus (\mathbf{1},2)\) and \(q^c \sim (\mathbf{2},4) \oplus (\mathbf{1},2)\). 
Employing the results of Ref.~\cite{Novichkov:2021evw} summarised in~\cref{tab:decomp},
one finds the decompositions
\(Q \leadsto \mathbf{1}_1 \oplus \mathbf{1}_2 \oplus \mathbf{1}_2\) and
\(q^c \leadsto \mathbf{1}_2 \oplus 1_0 \oplus \mathbf{1}_2\)
under the residual \(\mathbb{Z}_3\) symmetry group at the cusp.
In an appropriate basis, if one considers that sub-blocks of \(M\) are on the same footing, the mass matrix will schematically have the structure
\begin{equation}\label{eq:struct}
    M \sim \begin{pmatrix}
        \begin{tabular}{c c | c}
        1 & \(\epsilon\) & 1 \\
        \(\epsilon\) & \(\epsilon^2\) & \(\epsilon\) \\ \hline
        1 & \(\epsilon\) & 1 
    \end{tabular}
    \end{pmatrix}\,,
   \quad \text{instead of} \quad
   \begin{pmatrix}
        \begin{tabular}{c c | c}
        1 & \(\epsilon\) & 1 \\
        \(\epsilon\) & \(\epsilon^2\) & \(\epsilon\) \\ \hline
        \(\epsilon\) & \(\epsilon^2\) & \(\epsilon\)
    \end{tabular}
    \end{pmatrix}
\end{equation}
as one may expect from the residual decomposition analysis.%
\footnote{Working out this particular example, one additionally finds that \(M_{12} = M_{21} = 0\) even though these entries can be as large as \(\mathcal{O}(\epsilon)\).}
In particular, \(M_{33}\) is populated by a modular singlet \(Y^{(4)}_\mathbf{1}\), whose norm, we argue, should not play a role in producing fermion mass hierarchies in a bottom-up approach.
In other words, even though \(Y^{(4)}_\mathbf{1}\) vanishes at the cusp, one is free to normalise it as \(Y^{(4)}_\mathbf{1} = 1\) elsewhere. 
Similarly, the doublet \(Y^{(6)}_\mathbf{2}\) which, following the definition given in~\mbox{\cref{app:mod_forms}}, reads \(Y^{(6)}_\mathbf{2} \sim (\epsilon^2,\epsilon)\) in an appropriate basis, may be normalised to read \(Y^{(6)}_\mathbf{2} \sim (\epsilon,1)\) in the vicinity of the cusp.
Nevertheless, we stress that the arbitrariness in these normalisations cannot change the ratios or relative suppressions between entries of modular multiplets. These are what one can reliably use to justify mass hierarchies.

\subsubsection{Minimal quark mass matrices}
\label{sec:minimal}

%%%%%%%%%%%%%%%%%%%%%%%%%%%%%%
\begin{table}[t]
  \centering
  \begin{tabular}{lcccccc}
    \toprule
    \(S_4'\) irrep \(\mathbf{r}\)&&\(\mathbf{1},\mathbf{1}',\mathbf{\hat{1}},\mathbf{\hat{1}}'\) & & \(\mathbf{2},\mathbf{\hat{2}}\) & &  \(\mathbf{3},\mathbf{3}',\mathbf{\hat{3}},\mathbf{\hat{3}}'\) \\
    \midrule
    \(\mathbb{Z}_3\) decomposition&&\(\mathbf{1}_k\) & &  \(\mathbf{1}_{k+1} \oplus \mathbf{1}_{k+2}\) & & \(\mathbf{1}_k \oplus \mathbf{1}_{k+1} \oplus \mathbf{1}_{k+2}\)
    \\
    \bottomrule
  \end{tabular}
  \caption{Decompositions of \(\Gamma_4' \simeq S_4'\) weighted multiplets \((\mathbf{r},k)\) under the residual symmetry group \(\mathbb{Z}_3^{ST}\) at the cusp~\cite{Novichkov:2021evw}. Subscripts are understood modulo 3.}
  \label{tab:decomp}
\end{table}
%%%%%%%%%%%%%%%%%%%%%%%%%%%%%%

It should be clear that, once the above viewpoint is adopted, a fully hierarchical \(M\) cannot contain four or more sub-blocks. It follows that either \(Q\) or \(q^c\) (or both) must furnish some triplet representation(s) \(\mathbf{3}^*\) of the finite modular group \(S_4'\). From the outset, it is not immediately important which of the two must be a triplet, since the spectrum is insensitive to transposition, i.e.~to the exchange of irrep and weight assignments of \(Q\) and \(q^c\).
Under the residual \(\mathbb{Z}^{ST}_3\) group at the cusp, one has \((\mathbf{3}^*,k) \leadsto \mathbf{1}_k \oplus \mathbf{1}_{k+1} \oplus \mathbf{1}_{k+2}\)~\cite{Novichkov:2021evw}, as shown in~\cref{tab:decomp}.
For any weight, this decomposition spans all three \(\mathbb{Z}_3\) irreps. To avoid more than one large mass in the vicinity of the symmetric point, the partners of \(\mathbf{3}^*\) need to decompose into a direct sum of three copies of the same \(\mathbb{Z}_3\) irrep.
This automatically precludes using doublet or triplet irreps for these fields, cf.~\cref{tab:decomp}.
Therefore, if (say) \(Q \sim (\mathbf{3}^*,k)\), then \(q^c \sim \bigoplus_\beta (\mathbf{1}_\beta,k_\beta )\) with all \(k_\beta\) equal modulo 3.
As such, the desired mass matrix can only originate from modular form triplets of weight \(k_Y + 3n\), with \(n \in \mathbb{Z}\).

We proceed in a systematical way and restrict our attention to mass matrices that i) do not lead to massless quarks, by imposing \(\det M \neq 0\), and that ii) involve {\it at most} 4 parameters in the corresponding sector (excluding \(\tau\)).
Note that in the presence of a gCP symmetry, this maximum number of real parameters (4 in each sector and 2 from \(\tau\)) already matches the number of quark observables (6 masses, 3 angles, and 1 CPV phase).
To list all viable matrices, one starts by counting the number of linearly independent \(S_4'\) triplet modular forms available at each weight. This counting is given in~\cref{tab:YukawaFormsCounting}, with each sub-table corresponding to a different weight modulo 3.

%%%%%%%%%%%%%%%%%%%%%%%%%%%%%%
\begin{table}[t]
  \centering
  \begin{tabular}{lcc}
    \toprule
         &  \(\mathbf{3}\) (\(\mathbf{\hat{3}'}\)) & \(\mathbf{3'}\) (\(\mathbf{\hat{3}}\))  \\
         \midrule
          \(k=1\)   & 0  & \color{blue}{1}  
          \\
          \(k=4\)   & \color{blue}{1}  & \color{blue}{1} 
          \\
          \(k=7\)   & \color{teal}{2}  & \color{teal}{2}  
          \\
          \(k=10\)   & \color{teal}{2}  & \color{red}{3} 
          \\
    \bottomrule
  \end{tabular} \quad
    \begin{tabular}{lcc}
    \toprule
         &  \(\mathbf{3}\) (\(\mathbf{\hat{3}'}\)) & \(\mathbf{3'}\) (\(\mathbf{\hat{3}}\)) \\
         \midrule
          \(k=2\)   & 0  & \color{blue}{1}  
          \\
          \(k=5\)  & \color{blue}{1}  & \color{teal}{2} 
          \\
          \(k=8\)   & \color{teal}{2}  & \color{teal}{2}  
          \\
          \(k=11\)   & \color{red}{3}  & \color{red}{3}  
          \\
    \bottomrule
  \end{tabular} \quad
    \begin{tabular}{lcc}
    \toprule
         &  \(\mathbf{3}\) (\(\mathbf{\hat{3}'}\)) & \(\mathbf{3'}\) (\(\mathbf{\hat{3}}\)) \\
         \midrule
          \(k=3\textcolor{white}{0}\)   & \color{blue}{1}  & \color{blue}{1}  
          \\
          \(k=6\textcolor{white}{0}\)   & \color{blue}{1}  & \color{teal}{2}  
          \\
          \(k=9\textcolor{white}{0}\)   & \color{teal}{2}  & \color{red}{3}  
          \\
          \(k=12\)   & \color{red}{3}  & \color{red}{3}  
          \\
    \bottomrule
  \end{tabular}
  \caption{
  Counting of linearly independent \(S_4'\) modular forms for a given weight \(k\) and triplet representation. 
  For even (odd) weights, the counting refers to the triplet representation shown outside (within) the parentheses.
  }
  \label{tab:YukawaFormsCounting}
\end{table}
%%%%%%%%%%%%%%%%%%%%%%%%%%%%%%

Viable minimal matrices involve at most 4 independent triplet modular forms, for 3 different \((\mathbf{r},k)\) pairs chosen from the same sub-table. Therefore, any \((\mathbf{r},k)\) pair for which there are ``\(\color{red}{3}\)'' or more independent forms is excluded.
The only possibilities correspond to either selecting three ``\(\color{blue}{1}\)'' entries (2 possible three-parameter matrices), or two ``\(\color{blue}{1}\)'' entries together with a single ``\(\color{teal}{2}\)'' entry (18 four-parameter matrices). However, not all of these 20 matrices are viable, since some of the modular forms turn out to be proportional to each other across weights, leading to a massless quark.
Indeed, the fact that \(Y_\mathbf{\hat{3}}^{(1)} \propto Y_\mathbf{3}^{(4)}\) and \(Y_\mathbf{\hat{3}}^{(3)} \propto Y_\mathbf{3}^{(6)}\) means that all three-parameter matrices have zero determinant. 
We are left with only 6 viable four-parameter matrices \(M_i\) (\(i=1,\ldots,6\)), defined in~\cref{tab:kr} via the weights and irreps of the forms entering them.

%%%%%%%%%%%%%%%%%%%%%%%%%%%%%%
\begin{table}[t]
  \centering
  \begin{tabular}{lcccccc}
    \toprule
     & \(M_1\) & \(M_2\) & \(M_3\) & \(M_4\) & \(M_5\) & \(M_6\) \\
    \midrule
    \((\mathbf{r}_1,k_1)\) & 
    \((\mathbf{\hat{3}},1)\) &
    \((\mathbf{3},4)\) &
    \((\mathbf{3'},2)\) &
    \((\mathbf{3'},2)\) &
    \((\mathbf{\hat{3}'},3)\) &
    \((\mathbf{\hat{3}'},3)\)
    \\
    \((\mathbf{r}_2,k_2)\) &
    \((\mathbf{3'},4)\) &
    \((\mathbf{3'},4)\) &
    \((\mathbf{\hat{3}'},5)\) &
    \((\mathbf{\hat{3}'},5)\) &
    \((\mathbf{\hat{3}},3)\) &
    \((\mathbf{3},6)\)
    \\
    \((\mathbf{r}_3,k_3)\) &
    \((\mathbf{\hat{3}'},7)\) &
    \((\mathbf{\hat{3}'},7)\) &
    \((\mathbf{\hat{3}},5)\) &
    \((\mathbf{3},8)\) &
    \((\mathbf{3'},6)\) &
    \((\mathbf{3'},6)\)
    \\
    \bottomrule
  \end{tabular}
  \caption{Modular weights and representations for the minimal \(S_4'\) quark mass matrices viable in the vicinity of the cusp.}
  \label{tab:kr}
\end{table}
%%%%%%%%%%%%%%%%%%%%%%%%%%%%%%

Being more explicit, note that e.g.~a variant of the matrix \(M_1\) with \((\mathbf{r}_2, k_2) = (\mathbf{3},4)\) or a variant of \(M_6\) with \((\mathbf{r}_1, k_1) = (\mathbf{\hat{3}},3)\)
are excluded since \(Y_\mathbf{3}^{(4)} \propto Y_\mathbf{\hat{3}}^{(1)}\) and
\(Y_\mathbf{3}^{(6)}  \propto Y_\mathbf{\mathbf{\hat{3}}}^{(3)}\), as previously indicated,
leading to a zero eigenvalue.
Instead, one can exclude the variant of \(M_1\) with \((\mathbf{r}_3, k_3) = (\mathbf{\hat{3}},7)\), since it turns out that \(Y_{\mathbf{\hat{3}},1}^{(7)} \propto Y_\mathbf{3'}^{(4)} - 3\frac{Y_{\mathbf{1}}^{(6)}}{Y_{\mathbf{\hat{1}'}}^{(3)}} \, Y_{\mathbf{\hat{3}}}^{(1)} \), while \(Y_{\mathbf{\hat{3}},2}^{(7)} \propto Y_\mathbf{3'}^{(4)}\). A similar situation happens for, e.g., the variant of \(M_4\) with \((\mathbf{r}_3, k_3) = (\mathbf{3'},8)\), where \(Y_{\mathbf{3'},1}^{(8)} \propto Y_\mathbf{\hat{3}'}^{(5)} + \frac{24}{17} \frac{Y_{\mathbf{1}}^{(6)}}{Y_{\mathbf{\hat{1}'}}^{(3)}}\,Y_\mathbf{3'}^{(2)}\)
and \(Y_{\mathbf{3'},2}^{(8)} \propto Y_\mathbf{\hat{3}'}^{(5)}\).

\vskip 2mm

The matrices \(M_i\) (\(i=1,\ldots,6\)), that we consider in what follows, are the minimal mass matrices, \emph{involving at most four superpotential parameters and no massless fermions}, where the quark mass hierarchies may stem from the proximity of \(\tau\) to the cusp, assuming our choice of modular form normalisation. 
In the limit where \(\tau\) is brought to the symmetric point (\(\epsilon\to 0\)), all of them lead to a single massive quark.
Furthermore, since \(Y^{(1)}_{\mathbf{\hat{3}}} \propto Y^{(4)}_{\mathbf{3}}\) and \(Y^{(3)}_{\mathbf{\hat{3}}} \propto Y^{(6)}_{\mathbf{3}}\), \(M_1\) and \(M_2\) as well as \(M_5\) and \(M_6\) are effectively the same matrices and need not be considered separately.
In~\cref{sec:matrices}, we derive approximate analytical expressions for fermion mass ratios for each of the \(M_i\).

%%%%%%%%%%%%%%%%%%%%%%%
\subsection{Assignments, transposition and gCP}
\label{sec:notes}
%%%%%%%%%%%%%%%%%%%%%%%

Before proceeding, note that the up and down sectors are connected by the quark doublets \(Q\). Therefore, it is not obvious whether one may choose independently \(M_u\) and \(M_d\) from the above list of 6 matrices, i.e.~whether one may find modular \(S_4'\) assignments for \(Q\), \(u^c\) and \(d^c\) leading to every possible matrix pair. 

Let us consider separately the cases \(Q \sim \mathbf{3}^*\) and \(q^c \sim \mathbf{3}^*\) (\(q^c = u^c,d^c\)). 
If one takes \(Q \sim \mathbf{3}^*\), there is still freedom to adjust the singlets \(u^c_\beta\) and \(d^c_\beta\) independently so that each product \(Q \, u^c_\beta\) and \(Q \, d^c_\beta\) carries the desired weight and furnishes the desired triplet representation.%
\footnote{Indeed, this is an underconstrained problem, as there are classes of assignments that lead to the same quark mass matrix. For example, the choices \(Q \sim  (\mathbf{3},k)\), \(q^c \sim (\mathbf{\hat{1}'},-k+1) \oplus (\mathbf{1'},-k+4) \oplus (\mathbf{\hat{1}},-k+7)\) and \(Q \sim (\mathbf{3}',k)\), \(q^c \sim (\mathbf{\hat{1}},-k+1) \oplus (\mathbf{1},-k+4) \oplus (\mathbf{\hat{1}'},-k+7)\) both generate \(M_1\).}
As such, any choice \((M_u,M_d) = (M_i,M_j)\) with \(i,j=1,\ldots,6\) is possible
in principle --- their compatibility with quark mixing will be discussed in~\cref{sec:results}.
The same is not true for the case where \(u^c\) and \(d^c\) are triplets, as shown below. 

In the case where \(q^c\) are triplets, one has \(Q \sim \bigoplus_\alpha (\mathbf{1}_\alpha,k_\alpha)\) with all \(k_\alpha\) equal modulo 3.%
\footnote{The ordering of singlets is unphysical: any permutation corresponds to a weak basis choice affecting both sectors simultaneously and thus cannot impact quark mixing.}
These 1-dimensional irreps are shared by \(M_u\) and \(M_d\). It follows that the modular forms that feature in \(M_u\) and \(M_d\) must differ by a common weight, \(k_u^c - k_d^c\), and by a single 1-dimensional irrep factor.
In other words, if \(M_u\) is built from modular forms furnishing the representations \((\mathbf{r}_1, \mathbf{r}_2, \mathbf{r}_3)\), those entering \(M_d\) must correspond to \((\mathbf{r}_1, \mathbf{r}_2, \mathbf{r}_3) \otimes \mathbf{1}^*\), where \(\mathbf{1}^*\) is some 1-dimensional irrep. This reduces the possible pairs \((M_u,M_d)\) in the \(q^c \sim \mathbf{3}^*\) case to
\begin{equation}
\begin{aligned}
&(M_1^T,M_4^T)\,, \quad
(M_2^T,M_5^T)\,, \quad
(M_3^T,M_6^T)\,, \quad \\
&(M_4^T,M_1^T)\,, \quad
(M_5^T,M_2^T)\,, \quad
(M_6^T,M_3^T)\,, \quad\text{and}\quad (M_i^T,M_i^T)\text{ with }i=1,\ldots,6\,.
\end{aligned}
\end{equation}
Note that the relevant mass matrices must be transposed with respect to the previous case of \(Q \sim \mathbf{3}^*\).

Finally let us comment on the number of real parameters brought about by these models in the presence or absence of a gCP symmetry.
Each mass matrix will be a function of four superpotential parameters \(\alpha_i\) (\(i=1,\ldots,4\)) which can be complex, in general.
Schematically (\(q=u,d\)),
\begin{equation}\label{eq:notation1}
    M_q \sim \begin{pmatrix}
|&|&|\\
        \alpha_1 Y_1 & \alpha_2 Y_2 & \alpha_3 Y_3 + \alpha_4 Y_4\\
|&|&|
    \end{pmatrix} \text{ or its transpose}\,,
\end{equation}
where the \(Y_i\) denote the triplet modular forms after a Clebsch-Gordan rearrangement.
Further details are given in the next section, see~\cref{eq:matricesform}.
Imposing gCP, all \(\alpha_i\) are made real%
\footnote{%
In our case, CP invariance implies the reality of superpotential parameters,
as we are considering a symmetric basis for the generators of \(S_4'\) as well as real Clebsch-Gordan coefficients~\cite{Novichkov:2019sqv,Novichkov:2020eep}.}
 and the number of real parameter matches the number of observables, as previously commented: \((4 \times 2) + 2 = 10\), independently of whether \(Q\) or \(q^c\) is a triplet.
Lifting the requirement of gCP in the case \(Q \sim \mathbf{3}^*\) generically leads to one extra physical phase in each matrix, since the \(u^c_\beta\) and \(d^c_\beta\) can be rephased independently to absorb all \(\arg(\alpha_{1,2,3})\). Therefore, in each sector we are left with \(\alpha_4\) as the only complex parameter, for a total of \((5 \times 2) + 2 = 12\) degrees of freedom in the quark model.
Instead, the absence of gCP in the \(q^c \sim \mathbf{3}^*\) (transpose) case 
allows, in general, for \((5+7) + 2 = 14\) real degrees of freedom, since one 
may only absorb the phases \(\arg(\alpha_{1,2,3})\) in one of the sectors, 
via the rephasing of the \(Q_\alpha\), and a global phase in the other sector, 
via \(q^c\) rephasing. 
In light of this proliferation of free real parameters, we will not consider the latter class of models --- transpose case without gCP --- in what follows.

%%%%%%%%%%%%%%%%%%%%%%%%%%%%%%%%%%%%%%%%%%%%%%
\section{Analytical results for mass matrices}
\label{sec:matrices}
%%%%%%%%%%%%%%%%%%%%%%%%%%%%%%%%%%%%%%%%%%%%%%

In this section we derive approximate analytical expressions for the fermion mass ratios for each of the minimal \(S'_4\) quark matrices identified in the previous section.
The superpotential of interest reads (\(q=u,d\)):
\begin{equation}
\begin{aligned}
    W_q &= \alpha_1 \left(Y^{(k_1)}_{\mathbf{r}_1} \,Q\, q_1^c\right)_\mathbf{1} H_q
+  \alpha_2 \left(Y^{(k_2)}_{\mathbf{r}_2} \,Q\, q_2^c\right)_\mathbf{1} H_q \\
&+  \alpha_3 \left(Y^{(k_3)}_{\mathbf{r}_3,1} \,Q\, q_3^c\right)_\mathbf{1} H_q
+  \alpha_4 \left(Y^{(k_3)}_{\mathbf{r}_3,2} \,Q\, q_3^c\right)_\mathbf{1} H_q
\,,
\end{aligned}
\end{equation}
where we set \(\alpha_{1,2,3}\) real and non-negative without loss of generality
(since only \(|\alpha_{1,2,3}|^2\) will enter in the expressions for \(M_q M^\dagger_q\) whose eigenvalues are the corresponding quark masses squared and whose diagonalising unitary matrix enters the expression for the CKM quark mixing matrix),
while \(\alpha_4 \in \mathbb{C}\) in general, covering the no-gCP case.
The weights and representations of the modular forms entering \(W_q\) have been summarised in~\cref{tab:kr} for each of the six quark mass matrices \(M_i\) (\(i=1,\ldots, 6\)) under consideration.
Before taking into account the canonical normalisation of the fields (see also~\cref{sec:results}),
these matrices take the form
\begin{equation}
\begin{aligned}\label{eq:matricesform}
M_i &= v_q \left[
\frac{\alpha_1}{\sqrt{3}} \begin{pmatrix} y_1 & 0 & 0 \\ y_3 & 0& 0\\ y_2 & 0& 0 \end{pmatrix}_{Y^{(k_1)}_{\mathbf{r}_1}} 
+
\frac{\alpha_2}{\sqrt{3}} \begin{pmatrix} 0 & y_1& 0\\ 0 & y_3 & 0 \\  0 & y_2& 0 \end{pmatrix}_{Y^{(k_2)}_{\mathbf{r}_2}}
\right.
\\
&\qquad +
\left.
\frac{\alpha_3}{\sqrt{3}} \begin{pmatrix}  0 & 0& y_1\\ 0 & 0& y_3 \\ 0 & 0 & y_2 \end{pmatrix}_{Y^{(k_3)}_{\mathbf{r}_3,1}} 
+
\frac{\alpha_4}{\sqrt{3}} \begin{pmatrix} 0 & 0& y_1\\ 0 & 0& y_3 \\ 0 & 0 & y_2 \end{pmatrix}_{Y^{(k_3)}_{\mathbf{r}_3,2}} 
\right]\,,
\end{aligned}
\end{equation}
in the left-right convention.
In what follows, we expand each of these matrices in leading order in \(|\epsilon|\) and obtain approximate expressions for \(q\)-quark masses and quark mass ratios.
Masses are defined by the ordering \(m_1 \ll m_2 \ll m_3\), which may require appropriate permutations in the diagonalisation of the Hermitian products \(M_q M_q^\dagger\).

Note that these results apply not just within the quark sector but to any fermionic sector of the theory where hierarchical structures may be required (e.g.~the charged-lepton sector).
We further assume an appropriate rotation of left-handed fields and the rephasing of right-handed fields, in order to move to a common ``\(ST\)-diagonal'' weak basis where the power structure in \(|\epsilon|\) is apparent, see also~\cref{app:mod_forms}. Here, we consider \(\epsilon \in \mathbb{C}\) as defined in~\cref{eq:epsilon} (the absolute value is no longer implied).

\subsection*{Mass matrix \texorpdfstring{\(M_1\)}{M1}}
Keeping only the leading term in \(|\epsilon|\) element-wise, this matrix is given by the product
\begin{equation}
\begin{aligned}
M_1 &\simeq v_q \,\ta_1
\begin{pmatrix}
\dfrac{\epsilon}{\sqrt{3}} & - \sqrt{3}\,\epsilon 
    & -\dfrac{\epsilon}{\sqrt{3}} \left(7-\dfrac{\ta_4}{\ta_3}\right)
\\[4mm]
- \dfrac{\epsilon^2}{6} & \dfrac{7\epsilon^2}{6}
    & - \dfrac{\epsilon^2}{6} \left(49 + \dfrac{\ta_4}{\ta_3}\right)
\\[4mm]
1 & 1 & 1 + \dfrac{\ta_4}{\ta_3}
\end{pmatrix} \cdot
\begin{pmatrix}
    1 & 0 & 0 \\
    0 & \ta_2 & 0 \\
    0 & 0 & \ta_3
\end{pmatrix}
\,,
\end{aligned}
\end{equation}
after accounting for the canonical normalisation of the fields, with
\begin{equation}
    \begin{aligned}
    \frac{\ta_1}{\sqrt{2 \im \tau}}\ &\,=\, 
    \left(\sqrt{3}-1\right)|\theta|^2 \alpha_1 
    \,\simeq\, 0.73 |\theta|^2 \alpha_1 \,,\\
    %%%%%%%%%%%%%%%%%%%%%%%%%%%%%%%%%%%%%%%%%%%%%%%%%%%%%%%%%%%
    \frac{\ta_2}{(2 \im \tau)^{3/2}} &\,=\, 
    \left(9-5\sqrt{3}\right)|\theta|^6
    \frac{\alpha_2}{\alpha_1} 
    \,\simeq\, 0.34 |\theta|^6\frac{\alpha_2}{\alpha_1}\,,\\
    %%%%%%%%%%%%%%%%%%%%%%%%%%%%%%%%%%%%%%%%%%%%%%%%%%%%%%%%%%%
    \frac{\ta_3}{({2 \im \tau})^3} &\,=\, 
    18\sqrt{\frac{2}{37}}\left(26-15\sqrt{3}\right)|\theta|^{12}\frac{\alpha_3}{\alpha_1} 
    \,\simeq\, 0.08|\theta|^{12}\frac{\alpha_3}{\alpha_1}\,,\\
    %%%%%%%%%%%%%%%%%%%%%%%%%%%%%%%%%%%%%%%%%%%%%%%%%%%%%%%%%%%
    \frac{\ta_4}{({2 \im \tau})^3} &\,=\, 
    18\sqrt{2}\left(26-15\sqrt{3}\right)|\theta|^{12}\frac{\alpha_4}{\alpha_1} 
    \,\simeq\, 0.49|\theta|^{12}\frac{\alpha_4}{\alpha_1}\,.\\
    \end{aligned}
\end{equation}
The ensuing masses and mass ratios are given by
\begin{align} \label{eq:M1ratios1}
 m_3 &\,\simeq\, 
v_q\, \ta_1 \,\sqrt{1+ \ta_2^2 + |\ta_3+\ta_4|^2}\,,
 \\[2mm]
\frac{m_2}{m_3} &\,\simeq\, \frac{4}{\sqrt{3}}
\frac{\sqrt{\left(1+|\ta_3-\ta_4|^2\right)\ta_2^2+4\ta_3^2}}%
{1+\ta_2^2+ |\ta_3+\ta_4|^2}
|\epsilon|
\,,
\label{eq:M1ratios2} \\
\frac{m_1}{m_3} &\,\simeq\,
\frac{32}{3}
\frac{\ta_2 \ta_3}%
{\sqrt{1+\ta_2^2+ |\ta_3+\ta_4|^2}
\sqrt{\left(1+|\ta_3-\ta_4|^2\right)\ta_2^2+4\ta_3^2}}
|\epsilon|^2
\,, \label{eq:M1ratios3}
\end{align}
while for the determinant one obtains
\begin{align}
|\det M_1\,| &\simeq  \frac{128}{3\sqrt{3}}v_q^3 \ta_1^3 \ta_2 \ta_3\,|\epsilon|^3\,. 
\end{align}
In modular models with a single modulus, the small value of \(|\epsilon|\) is shared by both the up and down sectors. It is challenging to fit both up- and down-quark mass hierarchies using the same power structure, \(1:|\epsilon|:|\epsilon|^2\), in \(M_u\) and \(M_d\).
An additional suppression of quark mass ratios in one of the sectors may be arranged if, e.g., one of the superpotential constants is sufficiently larger than the others. Namely, the useful limit corresponds to  taking the constant which is absent from the determinant to be large.
Accordingly, in the limit \(|\ta_4| \gg \ta_2, \ta_3\) one finds
\begin{equation}
    m_3 \simeq v_q \ta_1 |\ta_4|\,,  \qquad
    \frac{m_2}{m_3} \simeq \frac{4}{\sqrt{3}} \ta_2 \left| \frac{\epsilon}{\ta_4}  \right| \,,  \qquad 
    \frac{m_1}{m_3} \simeq \frac{32}{3} \ta_3 \left| \frac{\epsilon }{\ta_4}\right|^2  \,,
\end{equation}
illustrating how a different hierarchy may arise in both sectors (see also~\cite{Petcov:2022fjf}).

\subsection*{Mass matrix \texorpdfstring{\(M_2\)}{M2}}
The results for \(M_2\) coincide with those for \(M_1\), provided one redefines the \(\ta_{1,\ldots,4}\) as
\begin{equation}
    \begin{aligned}
    \frac{\ta_1}{{(2 \im \tau)^2}} &\,=\, 
    6\left(7-4\sqrt{3}\right)|\theta|^8\alpha_1 
    \,\simeq\, 0.43 |\theta|^8\alpha_1 \,,\\
    %%%%%%%%%%%%%%%%%%%%%%%%%%%%%%%%%%%%%%%%%%%%%%%%%%%%%%%%%%%
    {\ta_2} &\,=\, 
    \frac{1}{\sqrt{3}}    \frac{\alpha_2}{\alpha_1} 
    \,\simeq\, 0.58 \frac{\alpha_2}{\alpha_1}\,,\\
    %%%%%%%%%%%%%%%%%%%%%%%%%%%%%%%%%%%%%%%%%%%%%%%%%%%%%%%%%%%
    \frac{\ta_3}{{(2 \im \tau)^{3/2}}} &\,=\, 
    3\sqrt{\frac{2}{37}}\left(3\sqrt{3}-5\right)|\theta|^6\frac{\alpha_3}{\alpha_1} 
    \,\simeq\, 0.14|\theta|^6\frac{\alpha_3}{\alpha_1}\,,\\
    %%%%%%%%%%%%%%%%%%%%%%%%%%%%%%%%%%%%%%%%%%%%%%%%%%%%%%%%%%%
    \frac{\ta_4}{{(2 \im \tau)^{3/2}}} &\,=\, 
    3\sqrt{2}\left(3\sqrt{3}-5\right)|\theta|^6\frac{\alpha_4}{\alpha_1} 
    \,\simeq\, 0.83|\theta|^6\frac{\alpha_4}{\alpha_1}\,.\\
    \end{aligned}
\end{equation}
This can be traced to the fact that the modular forms entering the second columns of \(M_1\) and \(M_2\) are proportional to each other, \(Y^{(4)}_{\mathbf{3}} \propto Y^{(1)}_{\mathbf{\hat{3}}}\), as previously indicated.

\subsection*{Mass matrix \texorpdfstring{\(M_3\)}{M3}}
This mass matrix has the approximate form
\begin{equation}
\begin{aligned}
M_3 &\simeq v_q \,\ta_1
\begin{pmatrix}
- \dfrac{\epsilon^2}{2}  & \dfrac{3\,\epsilon^2}{2}  & 
- \dfrac{\epsilon^2}{2}  \left(5-\dfrac{\ta_4}{\ta_3}\right)
\\[4mm]
1 & 1 & 1 - \dfrac{\ta_4}{\ta_3}\\[4mm]
\dfrac{\epsilon}{\sqrt{3}}  & 
-  \dfrac{\epsilon}{\sqrt{3}}  & 
-\dfrac{\epsilon}{\sqrt{3}}  \left(5+\dfrac{\ta_4}{\ta_3}\right)
\end{pmatrix}
\cdot
\begin{pmatrix}
    1 & 0 & 0 \\
    0 & \ta_2 & 0 \\
    0 & 0 & \ta_3
\end{pmatrix}
\,,
\end{aligned}
\end{equation}
with
\begin{equation}
    \begin{aligned}
    \frac{\ta_1}{2 \im \tau} &\,=\, 
    2\sqrt{14-8\sqrt{3}}\,|\theta|^4\alpha_1
    \,\simeq\, 0.76 |\theta|^4\alpha_1 \,,\\
    %%%%%%%%%%%%%%%%%%%%%%%%%%%%%%%%%%%%%%%%%%%%%%%%%%%%%%%%%%%
    \frac{\ta_2}{{(2 \im \tau)^{3/2}}} &\,=\, 
    \left(9-5\sqrt{3}\right)|\theta|^6 \frac{\alpha_2}{\alpha_1} 
    \,\simeq\, 0.34 |\theta|^6\frac{\alpha_2}{\alpha_1}\,,\\
    %%%%%%%%%%%%%%%%%%%%%%%%%%%%%%%%%%%%%%%%%%%%%%%%%%%%%%%%%%%
    \frac{\ta_3}{{(2 \im \tau)^{3/2}}} &\,=\, 
    \frac{3}{\sqrt{10}}\left(3\sqrt{3}-5\right)|\theta|^6\frac{\alpha_3}{\alpha_1} 
    \,\simeq\, 0.19|\theta|^6\frac{\alpha_3}{\alpha_1}\,,\\
    %%%%%%%%%%%%%%%%%%%%%%%%%%%%%%%%%%%%%%%%%%%%%%%%%%%%%%%%%%%
    \frac{\ta_4}{{(2 \im \tau)^{3/2}}} &\,=\, 
    3\left(3\sqrt{3}-5\right)|\theta|^6\frac{\alpha_4}{\alpha_1} 
    \,\simeq\, 0.59|\theta|^6\frac{\alpha_4}{\alpha_1}\,.\\
    \end{aligned}
\end{equation}
For the quark masses and mass ratios, one finds
(after applying the 2-3 permutation matrix on the 
diagonal matrix with the singular values of \(M_3\))
\begin{align}
 m_3 &\,\simeq\, 
v_q\, \ta_1 \,\sqrt{1+ \ta_2^2 + |\ta_3-\ta_4|^2}\,,
 \\[2mm]
\frac{m_2}{m_3} &\,\simeq\, \frac{2}{\sqrt{3}}
\frac{\sqrt{\left(1+|2\ta_3+\ta_4|^2\right)\ta_2^2+9\ta_3^2}}%
{1+\ta_2^2+ |\ta_3-\ta_4|^2}
|\epsilon|
\,,
\\
\frac{m_1}{m_3} &\,\simeq\,
\frac{8\, \ta_2 \ta_3}%
{\sqrt{1+\ta_2^2+ |\ta_3-\ta_4|^2}
\sqrt{\left(1+|2\ta_3+\ta_4|^2\right)\ta_2^2+9\ta_3^2}}
|\epsilon|^2
\,,
\end{align}
whereas for the determinant, 
\begin{align}
|\det M_3\,| &\simeq  \frac{16}{\sqrt{3}}v_q^3 \ta_1^3 \ta_2 \ta_3\,|\epsilon|^3\,,
\end{align}
which is independent of \(\ta_4\).
Accordingly, taking the limit \(|\ta_4| \gg \ta_2,\ta_3\) one finds
\begin{equation}
    m_3 \simeq v_q \ta_1 |\ta_4|\, ,  \qquad
    \frac{m_2}{m_3} \simeq \frac{2}{\sqrt{3}} \ta_2 \left| \frac{\epsilon}{\ta_4} \right| \, ,  \qquad 
    \frac{m_1}{m_3} \simeq 8 \ta_3 \left| \frac{\epsilon}{\ta_4} \right|^2  \, .
\end{equation}

\subsection*{Mass matrix \texorpdfstring{\(M_4\)}{M4}}
This mass matrix has the approximate form
\begin{equation}
\begin{aligned}
M_4 &\simeq v_q \,\ta_1
\begin{pmatrix}
- \dfrac{\epsilon^2}{2}  & \dfrac{3\,\epsilon^2}{2}  & 
- \dfrac{\epsilon^2}{2}  \left(5 +\dfrac{\ta_4}{\ta_3}\right)
\\[4mm]
1 & 1 & 1 + \dfrac{\ta_4}{\ta_3} \\[4mm]
\dfrac{\epsilon}{\sqrt{3}}  & 
-  \dfrac{\epsilon}{\sqrt{3}}  & 
-\dfrac{\epsilon}{\sqrt{3}}  \left(5-\dfrac{\ta_4}{\ta_3}\right)
\end{pmatrix}
\cdot
\begin{pmatrix}
    1 & 0 & 0 \\
    0 & \ta_2 & 0 \\
    0 & 0 & \ta_3
\end{pmatrix}
\,,
\end{aligned}
\end{equation}
with
\begin{equation}
    \begin{aligned}
    \frac{\ta_1}{2 \im \tau} &\,=\, 
    2\sqrt{14-8\sqrt{3}}\,|\theta|^4\alpha_1
    \,\simeq\, 0.76 |\theta|^4\alpha_1 \,,\\
    %%%%%%%%%%%%%%%%%%%%%%%%%%%%%%%%%%%%%%%%%%%%%%%%%%%%%%%%%%%
    \frac{\ta_2}{{(2 \im \tau)^{3/2}}} &\,=\, 
    \left(9-5\sqrt{3}\right)|\theta|^6 \frac{\alpha_2}{\alpha_1} 
    \,\simeq\, 0.34 |\theta|^6\frac{\alpha_2}{\alpha_1}\,,\\
    %%%%%%%%%%%%%%%%%%%%%%%%%%%%%%%%%%%%%%%%%%%%%%%%%%%%%%%%%%%
    \frac{\ta_3}{{(2 \im \tau)^3}} &\,=\, 
    \frac{18}{\sqrt{5}}\left(26-15\sqrt{3}\right)|\theta|^{12}\frac{\alpha_3}{\alpha_1} 
    \,\simeq\, 0.15|\theta|^{12}\frac{\alpha_3}{\alpha_1}\,,\\
    %%%%%%%%%%%%%%%%%%%%%%%%%%%%%%%%%%%%%%%%%%%%%%%%%%%%%%%%%%%
    \frac{\ta_4}{{(2 \im \tau)^3}} &\,=\, 
    18\sqrt{2}\left(26-15\sqrt{3}\right)|\theta|^{12}\frac{\alpha_4}{\alpha_1} 
    \,\simeq\, 0.49|\theta|^{12}\frac{\alpha_4}{\alpha_1}\,.\\
    \end{aligned}
\end{equation}
The quark masses and ratios follow:
\begin{align}
 m_3 &\,\simeq\, 
v_q\, \ta_1 \,\sqrt{1+ \ta_2^2 + |\ta_3+\ta_4|^2}\,,
 \\[2mm]
\frac{m_2}{m_3} &\,\simeq\, \frac{2}{\sqrt{3}}
\frac{\sqrt{\left(1+|2\ta_3-\ta_4|^2\right)\ta_2^2+9\ta_3^2}}%
{1+ \ta_2^2 + |\ta_3+\ta_4|^2}
|\epsilon|
\,,
\\
\frac{m_1}{m_3} &\,\simeq\,
\frac{8\, \ta_2 \ta_3}%
{\sqrt{1+ \ta_2^2 + |\ta_3+\ta_4|^2}
\sqrt{\left(1+|2\ta_3-\ta_4|^2\right)\ta_2^2+9\ta_3^2}}
|\epsilon|^2
\,,
\end{align}
and the determinant reads
\begin{align}
|\det M_4\,| &\simeq  \frac{16}{\sqrt{3}}v_q^3 \ta_1^3 \ta_2 \ta_3\,|\epsilon|^3\,. 
\end{align}
In the limit \(|\ta_4| \gg \ta_2, \ta_3\), one finds
\begin{equation}
    m_3 \simeq v_q \ta_1 |\ta_4|\, ,  \qquad
    \frac{m_2}{m_3} \simeq \frac{2}{\sqrt{3}} \ta_2 \left| \frac{\epsilon}{\ta_4} \right| \, ,  \qquad 
    \frac{m_1}{m_3} \simeq 8 \ta_3 \left| \frac{\epsilon}{\ta_4} \right|^2  \, .
\end{equation}

\subsection*{Mass matrix \texorpdfstring{\(M_5\)}{M5}}
This mass matrix has the approximate form
\begin{equation}
\begin{aligned}
M_5 &\simeq v_q \,\ta_1
\begin{pmatrix}
1 & 1 & 1 + \dfrac{\ta_4}{\ta_3} \\[4mm]
\sqrt{3}\,\epsilon  & 
-  \dfrac{\epsilon}{\sqrt{3}}  & 
-\sqrt{3}\,\epsilon \left(1-\dfrac{\ta_4}{\ta_3}\right)
\\[4mm]
- \dfrac{\epsilon^2}{2}  & 
  \dfrac{5\,\epsilon^2}{6}  & 
- \dfrac{\epsilon^2}{2}  \left(5+\dfrac{\ta_4}{\ta_3}\right)
\end{pmatrix}
\cdot
\begin{pmatrix}
    1 & 0 & 0 \\
    0 & \ta_2 & 0 \\
    0 & 0 & \ta_3
\end{pmatrix}
\,,
\end{aligned}
\end{equation}
with
\begin{equation}
    \begin{aligned}
    \frac{\ta_1}{{(2 \im \tau)^{3/2}}} &\,=\, 
    2\left(3\sqrt{3}-5\right)|\theta|^6\alpha_1
    \,\simeq\, 0.39 |\theta|^6\alpha_1 \,,\\
    %%%%%%%%%%%%%%%%%%%%%%%%%%%%%%%%%%%%%%%%%%%%%%%%%%%%%%%%%%%
    \ta_2 &\,=\, 
    \sqrt{\frac{3}{2}} \frac{\alpha_2}{\alpha_1} 
    \,\simeq\, 1.23 \frac{\alpha_2}{\alpha_1}\,,\\
    %%%%%%%%%%%%%%%%%%%%%%%%%%%%%%%%%%%%%%%%%%%%%%%%%%%%%%%%%%%
    \frac{\ta_3}{{(2 \im \tau)^{3/2}}} &\,=\, 
    \frac{6}{\sqrt{13}}\left(3\sqrt{3}-5\right)|\theta|^6\frac{\alpha_3}{\alpha_1} 
    \,\simeq\, 0.33|\theta|^6\frac{\alpha_3}{\alpha_1}\,,\\
    %%%%%%%%%%%%%%%%%%%%%%%%%%%%%%%%%%%%%%%%%%%%%%%%%%%%%%%%%%%
    \frac{\ta_4}{{(2 \im \tau)^{3/2}}} &\,=\, 
    3\left(3\sqrt{3}-5\right)|\theta|^6\frac{\alpha_4}{\alpha_1} 
    \,\simeq\, 0.59|\theta|^6\frac{\alpha_4}{\alpha_1}\,.\\
    \end{aligned}
\end{equation}
For quark masses and mass ratios, one finds
\begin{align}
 m_3 &\,\simeq\, 
v_q\, \ta_1 \,\sqrt{1+ \ta_2^2 + |\ta_3+\ta_4|^2}\,,
 \\[2mm]
\frac{m_2}{m_3} &\,\simeq\, \frac{2}{\sqrt{3}}
\frac{\sqrt{\left(4+|\ta_3-2\ta_4|^2\right)\ta_2^2+9\ta_3^2}}%
{1+\ta_2^2+ |\ta_3+\ta_4|^2}
|\epsilon|
\,,
\\
\frac{m_1}{m_3} &\,\simeq\,
\frac{8\, \ta_2 \ta_3}%
{\sqrt{1+\ta_2^2+ |\ta_3+\ta_4|^2}
\sqrt{\left(4+|\ta_3-2\ta_4|^2\right)\ta_2^2+9\ta_3^2}}
|\epsilon|^2
\,,
\end{align}
while for the determinant one obtains
\begin{align}
|\det M_5\,| &\simeq  \frac{16}{\sqrt{3}}v_q^3 \ta_1^3 \ta_2 \ta_3\,|\epsilon|^3\,. 
\end{align}
In the limit \(|\ta_4| \gg \ta_2, \ta_3\), one finds
\begin{equation}
    m_3 \simeq v_q \ta_1 |\ta_4|\, ,  \qquad
    \frac{m_2}{m_3} \simeq \frac{4}{\sqrt{3}} \ta_2 \left| \frac{\epsilon}{\ta_4} \right| \, ,  \qquad 
    \frac{m_1}{m_3} \simeq 4 \ta_3 \left| \frac{\epsilon}{\ta_4} \right|^2  \, .
\end{equation}

\subsection*{Mass matrix \texorpdfstring{\(M_6\)}{M6}}
The results for \(M_6\) coincide with those for \(M_5\), provided one redefines the \(\ta_{1,\ldots,4}\) as
\begin{equation}
    \begin{aligned}
    \frac{\ta_1}{{(2 \im \tau)^{3/2}}} &\,=\, 
    2\left(3\sqrt{3}-5\right)|\theta|^6\alpha_1 
    \,\simeq\, 0.39 |\theta|^6\alpha_1 \,,\\
    %%%%%%%%%%%%%%%%%%%%%%%%%%%%%%%%%%%%%%%%%%%%%%%%%%%%%%%%%%%
    \frac{\ta_2}{{(2 \im \tau)^{3/2}}} &\,=\, 
    3\left(9-5\sqrt{3}\right)|\theta|^6\frac{\alpha_2}{\alpha_1} 
    \,\simeq\, 1.02 |\theta|^6\frac{\alpha_2}{\alpha_1}\,,\\
    %%%%%%%%%%%%%%%%%%%%%%%%%%%%%%%%%%%%%%%%%%%%%%%%%%%%%%%%%%%
    \frac{\ta_3}{{(2 \im \tau)^{3/2}}} &\,=\, 
    \frac{6}{\sqrt{13}}\left(3\sqrt{3}-5\right)|\theta|^6\frac{\alpha_3}{\alpha_1} 
    \,\simeq\, 0.33|\theta|^6\frac{\alpha_3}{\alpha_1}\,,\\
    %%%%%%%%%%%%%%%%%%%%%%%%%%%%%%%%%%%%%%%%%%%%%%%%%%%%%%%%%%%
    \frac{\ta_4}{{(2 \im \tau)^{3/2}}} &\,=\, 
    3\left(3\sqrt{3}-5\right)|\theta|^6\frac{\alpha_4}{\alpha_1} 
    \,\simeq\, 0.59|\theta|^6\frac{\alpha_4}{\alpha_1}\,.\\
    \end{aligned}
\end{equation}
This can be traced to the fact that the modular forms entering the second columns of \(M_5\) and \(M_6\) are proportional to each other, \(Y^{(6)}_{\mathbf{3}} \propto Y^{(3)}_{\mathbf{\hat{3}}}\), as previously indicated.

\vskip 2mm
In the following section, we confront the above flavour textures with quark data. We analyse some benchmarks in more detail in~\cref{sec:natural}.

%%%%%%%%%%%%%%%%%%%%%%%%%%%%%%%%%%%%%%%%%%%%%%
\section{Numerical results}
\label{sec:results}
%%%%%%%%%%%%%%%%%%%%%%%%%%%%%%%%%%%%%%%%%%%%%%

In the preceding sections, we have identified the minimal structures that can be assigned to \(M_u\) and \(M_d\), each depending on 4 independent parameters. The resulting \(S_4'\) quark modular models may lead to hierarchical masses near the cusp and do not present more free parameters than observables when gCP is imposed. The next step is to verify numerically if any of these models can actually achieve a good fit of quark data, summarised in~\cref{tab:data}.
To quantify the goodness of fit, we consider the sum of one-dimensional \(\chi^2\) functions in a Gaussian approximation,
\begin{equation}
    \chi^2(\vec{p}) = \sum_{j=1}^8 \left(\frac{\Theta_j(\vec{p})-\Theta_j^{\text{b.f.}}}{\sigma_j}\right)^2\,, 
\end{equation}
which we seek to minimise,
and define \(N \sigma \equiv \sqrt{\chi^2}\).
Here, \(\Theta_j\) correspond to the values of the 8 observables in the right-hand side of~\cref{tab:data}, as predicted by the model under consideration, for a given set of parameters \(\vec{p}\), while  \(\Theta^\text{b.f.}_j\) denotes their high-energy best-fit values and \(\sigma_j\) are the corresponding \(1\sigma\) uncertainties.
Note that if a model successfully reproduces dimensionless observables, the mass scales in each sector can be easily recovered by a common rescaling of the corresponding superpotential parameters.

%
%%%%%%%%%%%%%%%%%%%%%%%%%%%%%%
\begin{table}[t]
  \centering
  \begin{tabular}{lc}
    \toprule
    Observable & Best-fit \(\pm \, 1 \sigma\) range\\
    \midrule
    \(y_u \,/\, 10^{-6}\) & \(2.92 \pm 1.81\) \\
    \(y_c \,/\, 10^{-3}\) & \(1.43 \pm 0.100\) \\
    \(y_t\) & \(0.534 \pm 0.0341\) \\
    \(y_d \,/\, 10^{-6}\) & \(4.81 \pm 1.06\) \\
    \(y_s \,/\, 10^{-5}\) & \(9.52 \pm 1.03\) \\
    \(y_b \,/\, 10^{-3}\) & \(6.95 \pm 0.175\) \\
    \bottomrule
  \end{tabular}
  \qquad
  \begin{tabular}{lc}
    \toprule
    Observable & Best-fit \(\pm \, 1 \sigma\) range\\
    \midrule
    \((m_u/m_c) \,/\, 10^{-3}\) & \(2.04 \pm 1.27\) \\
    \((m_c/m_t) \,/\, 10^{-3}\) & \(2.68 \pm 0.25\) \\
    \((m_d/m_s) \,/\, 10^{-2}\) & \(5.05 \pm 1.24\) \\
    \((m_s/m_b) \,/\, 10^{-2}\) & \(1.37 \pm 0.15\) \\
    \midrule
    \(\theta_{12} \,(\degree)\) & \(13.027\pm 0.0814\) \\
    \(\theta_{23} \,(\degree)\) & \(2.054\pm 0.384\) \\
    \(\theta_{13} \,(\degree)\) & \(0.1802 \pm 0.0281\) \\
    \midrule
    \(\delta_\text{CP} \,(\degree)\) & \(69.21\pm 6.19\) \\
    \bottomrule
  \end{tabular}
  \caption{Best-fit values and \(1\sigma\) ranges for the quark Yukawa couplings (left), quark mass ratios, mixing angles and CPV phase (right) at the high-energy scale of \(2 \times 10^6\) GeV for \(\tan\beta = 5\). The values and uncertainties of Yukawa couplings, mixing angles and the CPV phase are reproduced from~\cite{Okada:2020rjb} and obtained from Refs.~\cite{Antusch:2013jca,Bjorkeroth:2015ora}.}
  \label{tab:data}
\end{table}
%%%%%%%%%%%%%%%%%%%%%%%%%%%%%%

For the \(\chi^2\) minimisation procedure, 
\(\tau\) is scanned within the fundamental domain \(\mathcal{D}\) and kept close to the cusps, in the regions
\begin{align}
\left|\re \tau\right| > \frac{1}{2} - 0.025\,,\qquad
\im \tau < \sqrt{1- (\re\tau)^2}+ 0.05
\,,
\end{align}
in agreement with our goal.%
\footnote{
A model not fitting the quark data in these regions may yet be viable for other values of \(\tau \in \mathcal{D}\).}
For the left cusp, the region includes values of \(|\epsilon|\) as large as 0.1.
A priori, no constraints are imposed on the ranges of superpotential parameters, i.e.~they can be arbitrarily large or small. Hence, in this first step, we are not concerned with the particular normalisation of modular forms, which can be absorbed in the superpotential parameters.
Additionally, these parameters can absorb the effects of canonically normalising the fields, due to the assumed minimal-form Kähler potential,
\begin{equation} \label{eq:Kahler}
K(\tau,\bar\tau; \psi, \bar\psi)
\supset
-\Lambda_K^2 \log (2\im \tau) 
+ \sum_{\psi \,\in\, \{Q_\alpha,\, u^c_\beta,\, d^c_{\beta'}\}} \frac{\left|\psi\right|^2}{(2 \im \tau)^{k_\psi}}
\,,
\end{equation}
with \(\Lambda_K\) having mass dimension one. Namely, fields are scaled as \(\psi \to \sqrt{(2 \im \tau)^{k_\psi}} \psi\) to yield canonical kinetic terms. Mass matrices will be affected accordingly, with each contribution being scaled by a factor \(\sqrt{(2 \im \tau)^{k_Y}}\). Note that this factor depends only on the weight \(k_Y\) of the corresponding modular form and has been taken into account in the dictionary between the \(\alpha_i\) and \(\tilde\alpha_i\) (\(i=1,\ldots,4\)) of~\cref{sec:matrices}.
The impact of modular form and canonical field normalisations on fine-tuning will be discussed in~\cref{sec:hier}.

\vskip 2mm
In what follows, we present our numerical results, starting with the cases where gCP is imposed and \(\delta_\text{CP}\) is either absent from (\cref{sec:nophasefit}) or present in (\cref{sec:yesphasefit}) the fit.
Lifting gCP allows for 11-parameter phenomenological fits (\cref{sec:fits11}) or 12-parameter gCP-consistent fits (\cref{sec:fits11}). Finally, we consider fits with an additional modulus and gCP (\cref{sec:2tau}).
In the summary tables, \textcolor{red}{9+} indicates a value of \(\sqrt{\chi^2}>9\), while \textcolor{orange}{5+} refers to values in the range \(5 < \sqrt{\chi^2} < 9\). For fits with a minimum below \( 5 \sigma\), the value of \(\sqrt{\chi^2}\) is given explicitly.
As will be shown already in~\cref{sec:yesphasefit}, a fit of the 10 quark observables is not possible within the 10-parameter models. Moreover, even in the presence of additional parameters, fitting the quark data in the vicinity of the cusps is not guaranteed.

%%%%%%%%%%%%%%%%%%%%%%%
\subsection{Fits without \texorpdfstring{\(\delta_\text{CP}\)}{δCP} in the presence of gCP}
\label{sec:nophasefit}
%%%%%%%%%%%%%%%%%%%%%%%

We start by considering the minimal cases resulting from the imposition of a gCP symmetry, such that the complexity of the mass matrices may only originate from non-CP conserving values of the modulus, via the modular forms. It may be challenging to obtain sizeable CP violation in the vicinity of the cusps if \(\tau\) is the only source of CP violation (see also~\cite{Petcov:2022fjf}). Therefore, we first exclude the CPV phase from the list of fit observables, checking if the models can reproduce the quark masses and mixing angles.
As a result, fits near the left and right cusps are equivalent (see also the comment at the end of~\cref{sec:modframework}).

%%%%%%%%%%%%%%%%%%%%%%%%%%%%%%
\begin{table}[t]
\renewcommand{\arraystretch}{1.4}
  \centering
  \begin{subtable}[h]{0.48\textwidth}
  \centering
  \begin{tabular}{c|@{\,\,\,\,\,}cccc}
    \toprule
   \diagbox[innerwidth=1cm,height=1.2cm]{\footnotesize\(M_u\)}{\footnotesize\(M_d\)} & \(M_{1,2}\) & \(M_3\) & \(M_4\) & \(M_{5,6}\) \\ 
    \midrule
     \(M_{1,2}\) & 0.0 & \nine & \nine & \nine \\
     \(M_3\) & \nine & 0.0 & 0.0 & \nine \\
     \(M_4\) & \nine & 0.0 & 0.0 & \nine\\
     \(M_{5,6}\) & \nine & \nine & \nine & 0.0 \\
    \bottomrule
  \end{tabular}
  \caption{Models with \(Q\sim \mathbf{3}^*\)}
  \label{tab:gCPnophasea}
  \end{subtable}
  \begin{subtable}[h]{0.48\textwidth}
  \centering
  \begin{tabular}{c|@{\,\,\,\,\,}cccc}
    \toprule
   \diagbox[innerwidth=1cm,height=1.2cm]{\footnotesize\(M_u\)}{\footnotesize\(M_d\)}& \(M_{1,2}^T\) & \(M_3^T\) & \(M_4^T\) & \(M_{5,6}^T\) \\
    \midrule
     \(M_{1,2}^T\) & 0.0 & -- & 1.5 & 1.5 \\
     \(M_3^T\) & -- & 1.0 & -- & 1.0 \\
     \(M_4^T\) & 0.0 & -- & 1.0 & -- \\
     \(M_{5,6}^T\) & 0.0 & 1.5 & -- & 1.4 \\
    \bottomrule
  \end{tabular}
  \caption{Models with \(q^c\sim \mathbf{3}^*\)}
  \label{tab:gCPnophaseb}
  \end{subtable}
  \caption{
  Values of \(\sqrt{\chi^2}\) for quark fits of 10-parameter (gCP) models, without \(\delta_\text{CP}\), depending on which fields are taken as triplets of \(S_4'\). Note that some pairs \((M_u,M_d) = (M_i^T,M_j^T)\) are not allowed, as discussed in~\cref{sec:notes}.  }
  \label{tab:gCPnophase}
\end{table}
%%%%%%%%%%%%%%%%%%%%%%%%%%%%%%

Our results are summarised in~\cref{tab:gCPnophasea,tab:gCPnophaseb}
for the cases where the left- or right-handed quarks are \(S_4'\) triplets, respectively.
A value of \(N\sigma = \sqrt{\chi^2} = 0.0\) indicates that one can reproduce the central values for all the observables under consideration.
As discussed in~\cref{sec:minimal,sec:matrices}, the matrices \(M_1\) and \(M_2\) are equivalent up to a redefinition of superpotential parameters. Hence, these cases are grouped in the tables. The same goes for \(M_5\) and \(M_6\). Recall that in the context of \(q^c \sim \mathbf{3}^*\) models only some combinations \((M_i^T,M_j^T)\) are meaningful, c.f.~\cref{sec:notes}.

It is interesting to note that, among these 10-parameter models, some can easily fit quark masses and mixing while others cannot fit them at all (at \(9\sigma\) or worse). This can be understood by verifying that, for those cases, the CKM matrix approaches a non-viable form in the limit of vanishing \(|\epsilon|\). In other words, the diagonalisation of the mass matrices requires permutations which do not cancel in the product defining the CKM matrix. This is not so for \(q^c \sim \mathbf{3}^*\) models, since the transpositions lead to democratic-like \(M_q M_q^\dagger\) matrices for both sectors, in the same limit.

Finally, we have verified that the CPV character of \(\tau\) is not relevant for the goodness of fit in these scenarios. Namely, we have checked that fits of masses and mixing are still possible, with the same \(\chi^2\) values, for CP-conserving values of \(\tau\), i.e.~imposing either \(\re \tau = \pm 1/2\) or \(|\tau|^2 = 1\) for the same value of \(|\epsilon|\) (but possibly different values of the other parameters).

%%%%%%%%%%%%%%%%%%%%%%%
\subsection{Fits with \texorpdfstring{\(\delta_\text{CP}\)}{δCP} in the presence of gCP}
\label{sec:yesphasefit}
%%%%%%%%%%%%%%%%%%%%%%%

We now include the CPV phase in the list of fit observables.
To be more precise, and in the context of this section alone, we consider 
\(J_\text{CP} = (2.31 \pm 0.57)\times 10^{-5} \),
with
\(J_\text{CP} \equiv c_{12} c_{13}^2 c_{23} s_{12} s_{13} s_{23} \sin \delta_\text{CP}\)
(\(c_{ij} = \cos \theta_{ij}\), \(s_{ij} = \sin \theta_{ij}\)),
in the search for models with sufficient CP violation, 
since \(\delta_\text{CP}\) itself may not be a good indicator of the latter
when mixing angles driven to very small values.
By including a CPV observable in the fit, differences may arise depending on which cusp the modulus approaches. We thus analyse both cusps separately.

%%%%%%%%%%%%%%%%%%%%%%%%%%%%%%
\begin{table}[t]
\renewcommand{\arraystretch}{1.4}
  \centering  
  \begin{subtable}[h]{0.48\textwidth}
  \centering
  \begin{tabular}{c|@{\,\,\,\,\,}cccc}
    \toprule
   \diagbox[innerwidth=1cm,height=1.2cm]{\footnotesize\(M_u\)}{\footnotesize\(M_d\)} & \(M_{1,2}\) & \(M_3\) & \(M_4\) & \(M_{5,6}\) \\ 
    \midrule
     \(M_{1,2}\) & \(4.1\) & \nine & \nine & \nine \\
     \(M_3\) & \nine & \(4.1\) & \(4.1\) & \nine \\
     \(M_4\) & \nine & \(4.1\) & \(4.1\) & \nine\\
     \(M_{5,6}\) & \nine & \nine & \nine & \(4.1\) \\
    \bottomrule
  \end{tabular}
  \caption{Models with \(Q\sim \mathbf{3}^*\)}
  \label{tab:gCPphasea}
  \end{subtable}
  \begin{subtable}[h]{0.48\textwidth}
  \centering
  \begin{tabular}{c|@{\,\,\,\,\,}cccc}
    \toprule
   \diagbox[innerwidth=1cm,height=1.2cm]{\footnotesize\(M_u\)}{\footnotesize\(M_d\)}&
    \(M_{1,2}^T\) & \(M_3^T\) & \(M_4^T\) & \(M_{5,6}^T\) \\
    \midrule
     \(M_{1,2}^T\) & \(4.0\) & --    & \(4.3\) & \(4.3\) \\
     \(M_3^T\)     & --    & \(4.2\) & --    & \(4.1\) \\
     \(M_4^T\)     & \(4.0\) & --    & \(4.2\) & --    \\
     \(M_{5,6}^T\) & \(4.0\) & \(4.3\) & --    & \(4.3\) \\
    \bottomrule
  \end{tabular}
  \caption{Models with \(q^c\sim \mathbf{3}^*\)}
  \label{tab:gCPphaseb}
  \end{subtable}
  \caption{
  Values of \(\sqrt{\chi^2}\) for quark fits of 10-parameter (gCP) models, including \(\delta_\text{CP}\), depending on which fields are taken as triplets of \(S_4'\). All entries apply to both cusps.}
  \label{tab:gCPphase}
\end{table}
%%%%%%%%%%%%%%%%%%%%%%%%%%%%%%

Our results are summarised in~\cref{tab:gCPphasea,tab:gCPphaseb}
for the cases where the left- or right-handed quarks are \(S_4'\) triplets, respectively.
As one may expect, results are globally worse in the presence of an extra constraint. Moreover, these results suggest that the proximity to an enhanced symmetry point (either \(\omega\) or \(\omega +1\)) places too big a strain on the models for them to be able to comply with all quark data below the \(4 \sigma\) level.
By comparing~\cref{tab:gCPnophase,tab:gCPphase}, one sees that this failure is driven by the CPV observable.

\vskip 2mm

\begin{figure}[t]
\centering
\includegraphics[scale=0.8]{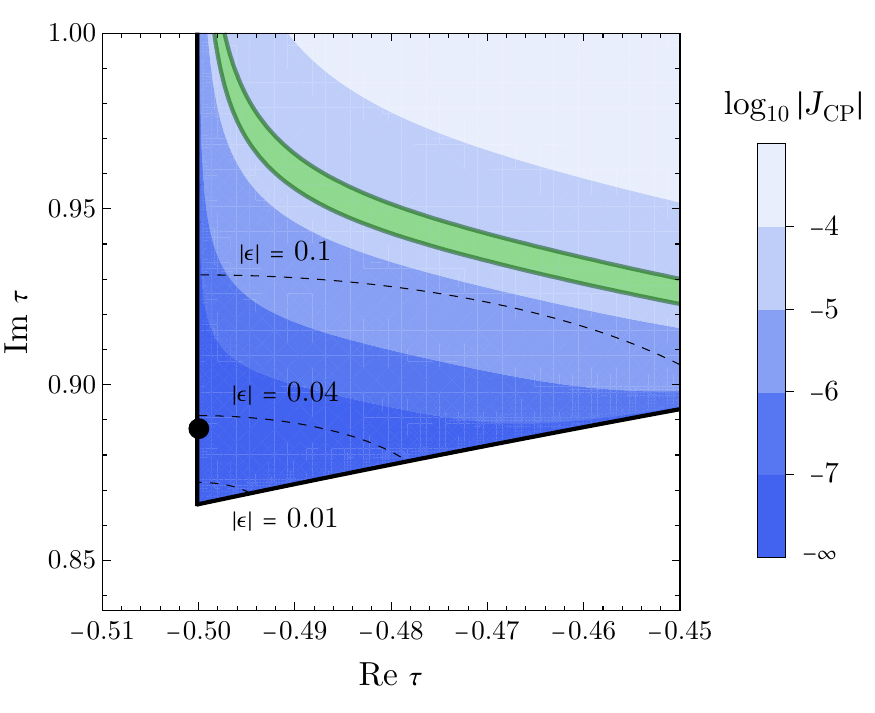}
\caption{Magnitude of CP violation for different values of the modulus \(\tau\) in the vicinity of the left cusp and for a fixed set of superpotential parameters. The model considered corresponds to \(M_{u,d} \sim M_1\) and can fit quark masses and mixing angles --- but not \(\delta_\text{CP}\) --- at the value of \(\tau\) marked by the black dot. The green band represents the 1\(\sigma\)-allowed range for the \(J_\text{CP}\) invariant.}
\label{fig:gCP}
\end{figure}

To illustrate the strain placed on the models by the addition of the CPV observable to the fit, consider~\cref{fig:gCP}, where a no-\(\delta_\text{CP}\) fit point for the \(M_{u,d} \sim M_1\) model (black dot) is shown in the \(\tau\) plane. One may quantify the magnitude of CP violation through the value of \(J_\text{CP}\) in this plane, by varying the value of \(\tau\) for this point while keeping the other parameters fixed. Note that this variation is done for illustrative purposes only, as it spoils the values of observables. One concludes that reaching the correct magnitude for \(J_\text{CP}\) (green band) calls for relatively large values of \(|\epsilon|\), i.e.~seems incompatible with the required closeness to the cusp.

%%%%%%%%%%%%%%%%%%%%%%%
\subsection{Fits with 11 parameters}
\label{sec:fits11}
%%%%%%%%%%%%%%%%%%%%%%%

So far, we have seen that the models of interest cannot comply with all quark data in the presence of a gCP symmetry. Lifting the assumption of gCP leads to new sources of complexity in the mass matrices, which may result in phenomenologically acceptable models.
As discussed at the end of~\cref{sec:notes}, these models in general will be described by 2 additional physical phases in the \(Q \sim \mathbf{3}^*\) case, namely the phases of the \(\alpha_4\) in the up and down sectors. We focus exclusively on this case in what follows, since the \(q^c \sim \mathbf{3}^*\) scenario brings about too many additional parameters (4 new phases). 

We start by following a \textit{phenomenological} approach, allowing only one of the \(\alpha_4\) parameters to be complex, in turn, for a total of 11 parameters.
Our results are summarised in~\cref{tab:nogCP}.
They are, in practice, independent of the cusp (left vs.~right) and of which sector (up vs.~down) contains the phenomenological phase.
By lifting gCP in this explicit way --- such that \(\tau\) is not the only source of CP violation --- models can be found which fit all the quark data. Nonetheless, note that the presence of more degrees of freedom than observables is not a sufficient condition for the model to be viable in the vicinity of the cusps (c.f.~also the discussion in~\cref{sec:nophasefit}).

%%%%%%%%%%%%%%%%%%%%%%%%%%%%%%
\begin{table}[t]
\renewcommand{\arraystretch}{1.4}
  \centering
  \begin{tabular}{c|@{\,\,\,\,\,}cccc}
    \toprule
   \diagbox[innerwidth=1cm,height=1.2cm]{\footnotesize\(M_u\)}{\footnotesize\(M_d\)} & \(M_{1,2}\) & \(M_3\) & \(M_4\) & \(M_{5,6}\) \\ 
    \midrule
     \(M_{1,2}\) & 0.0 & \nine & \nine & \nine \\
     \(M_3\) & \nine & 0.0 & 0.0 & \nine \\
     \(M_4\) & \nine & 0.0 & 0.0 & \nine\\
     \(M_{5,6}\) & \nine & \nine & \nine & 0.0 \\
    \bottomrule
  \end{tabular}
  \caption{ Values of \(\sqrt{\chi^2}\) for quark fits of 11-parameter (phenomenological) and 12-parameter (consistent) models with \(Q\sim \mathbf{3}^*\). These results apply to both cusps and are independent of which sector holds the new phase.}
  \label{tab:nogCP}
\end{table}
%%%%%%%%%%%%%%%%%%%%%%%%%%%%%%

%%%%%%%%%%%%%%%%%%%%%%%
\subsection{Fits with 12 parameters}
\label{sec:fits12}
%%%%%%%%%%%%%%%%%%%%%%%

Consistently allowing both \(\alpha_4\) to be complex results in a total of 12 parameters but leads to no apparent qualitative improvement with respect to the phenomenological case. In particular, the results as shown in~\cref{tab:nogCP} apply also to this general case.

It may be possible to reduce the number of degrees of freedom if, for instance, the value of the modulus is selected by a dynamical principle in a top-down approach. In Ref.~\cite{Novichkov:2022wvg}, simple modular-invariant supergravity-motivated potentials were considered and global CP-breaking minima were found in the vicinities of the cusps. The selected values of \(\tau\) follow a series,
\begin{equation}
\begin{aligned}
\tau \,\simeq\, \mp 0.484 + 0.884\, i,\,\, \mp 0.492 + 0.875\, i,\,\,\mp 0.495 + 0.872\, i,\,\,\ldots\,,\\ \text{ corresponding to }
|\epsilon(\tau)| \,\simeq \, 0.04,\,\, 0.02,\,\, 0.01,\,\, \ldots\,,
\label{eq:tdtau}
\end{aligned}
\end{equation}
which approaches the cusps. The goodness of fit is in general expected to decrease.
We find that for the first of the values in~\cref{eq:tdtau}, the viable fits remain acceptable, while this ceases to be the case as one gets closer to the cusps.
The corresponding results of these top-down inspired 10-parameter models are summarised in~\cref{tab:constraineda,tab:constrainedb}.

%%%%%%%%%%%%%%%%%%%%%%%%%%%%%%
\begin{table}[t]
\renewcommand{\arraystretch}{1.4}
  \centering  
  \begin{subtable}[h]{0.48\textwidth}
  \centering
  \begin{tabular}{c|@{\,\,\,\,\,}cccc}
    \toprule
   \diagbox[innerwidth=1cm,height=1.2cm]{\footnotesize\(M_u\)}{\footnotesize\(M_d\)} & \(M_{1,2}\) & \(M_3\) & \(M_4\) & \(M_{5,6}\) \\ 
    \midrule
     \(M_{1,2}\) & 0.1 & \nine & \nine & \nine \\
     \(M_3\) & \nine & 2.1 & 2.1 & \nine \\
     \(M_4\) & \nine & 2.1 & 2.1 & \nine\\
     \(M_{5,6}\) & \nine & \nine & \nine & 2.1 \\
    \bottomrule
  \end{tabular}
  \caption{\(\tau = \mp 0.484 + 0.884\,i\)}
  \label{tab:constraineda}
  \end{subtable}
  \begin{subtable}[h]{0.48\textwidth}
  \centering
  \begin{tabular}{c|@{\,\,\,\,\,}cccc}
    \toprule
   \diagbox[innerwidth=1cm,height=1.2cm]{\footnotesize\(M_u\)}{\footnotesize\(M_d\)} & \(M_{1,2}\) & \(M_3\) & \(M_4\) & \(M_{5,6}\) \\ 
    \midrule
     \(M_{1,2}\) & \five & \nine & \nine & \nine \\
     \(M_3\) & \nine & \five & \five & \nine \\
     \(M_4\) & \nine & \five & \five & \nine\\
     \(M_{5,6}\) & \nine & \nine & \nine & \five \\
    \bottomrule
  \end{tabular}
  \caption{\(\tau = \mp 0.492 + 0.875\,i\)}
  \label{tab:constrainedb}
  \end{subtable}
  \caption{
  Values of \(\sqrt{\chi^2}\) for quark fits of models in the absence of gCP, with \(Q \sim \mathbf{3}^*\) and fixing \(\tau\) to top-down selected values (10 free parameters). These results apply independently of the cusp considered.
  }
\end{table}
%%%%%%%%%%%%%%%%%%%%%%%%%%%%%%

%%%%%%%%%%%%%%%%%%%%%%%
\subsection{Fits with two moduli in the presence of gCP}
\label{sec:2tau}
%%%%%%%%%%%%%%%%%%%%%%%

Finally, we consider a phenomenologically-motivated analysis where we allow for two distinct moduli to co-exist, \(\tau_u\) coupling to up-type quarks and \(\tau_d\) coupling to down-type quarks. Such a situation may arise, for instance, in the context of symplectic modular invariance, see~\cite{Ding:2021iqp}.
Since now we have an extra complex parameter, gCP is imposed to keep the total number of parameters as small as possible: 12 for the minimal quark mass matrices identified.
The presence of an extra modulus allows for an independent source of CP violation in this scenario. Moreover, assuming that both moduli are near a cusp allows to decouple the explanation of quark mass hierarchies in the up and down sectors, which are now each controlled by their own small parameter, \(|\epsilon_u|\) and \(|\epsilon_d|\), respectively.%
\footnote{
Indeed, it is difficult to fit both quark sectors with a structure of the type \( 1 : |\epsilon| : |\epsilon|^2\), \(\mathcal{O}(1)\) coefficients, and a common value of \(|\epsilon|\).}

%%%%%%%%%%%%%%%%%%%%%%%%%%%%%%
\begin{table}[p]
\renewcommand{\arraystretch}{1.4}
  \centering  
  \begin{threeparttable}
  \begin{subtable}[h]{\textwidth}
  \centering
  \begin{tabular}{c|@{\,\,\,\,\,}cccc}
    \toprule
   \diagbox[innerwidth=1cm,height=1.2cm]{\footnotesize\(M_u\)}{\footnotesize\(M_d\)} & \(M_{1,2}\) & \(M_3\) & \(M_4\) & \(M_{5,6}\) \\ 
    \midrule
     \(M_{1,2}\) & \(0.0\) & \nine & \nine & \nine \\
     \(M_3\) & \nine & \(0.0\) & \(0.0\) & \nine \\
     \(M_4\) & \nine & \(0.0\) & \(0.0\) & \nine\\
     \(M_{5,6}\) & \nine & \nine & \nine & \(0.0\) \\
    \bottomrule
  \end{tabular}
  \caption{Models with \(Q\sim \mathbf{3}^*\)}
  \label{tab:twoa}
  \end{subtable}
  \vskip 5mm
  \begin{subtable}[h]{\textwidth}
  \centering
  \begin{tabular}{c|@{\,\,\,\,\,}cccc}
    \toprule
   \diagbox[innerwidth=1cm,height=1.2cm]{\footnotesize\(M_u\)}{\footnotesize\(M_d\)}&
    \(M_{1,2}^T\) & \(M_3^T\) & \(M_4^T\) & \(M_{5,6}^T\) \\
    \midrule
     \(M_{1,2}^T\) & \([0.0,0.4]\)& --    & \(\hphantom{^{***}}[0.0,2.8]\hphantom{^{***}}\) & \([0.0,0.7]\) \\[1mm]
     \(M_3^T\)     & --    & \(\hphantom{^{***}}[0.0,3.4]^*\hphantom{^{**}}\) & --    & \(\hphantom{^{***}}[1.1,3.4]^*\hphantom{^{**}}\) \\[1mm]
     \(M_4^T\)     & \(\hphantom{^{***}}[1.3,4.5]^{***}\) & --    & \([0.1,3.0]\) & --    \\[1mm]
     \(M_{5,6}^T\) & \([0.0,1.4]\) & \(\hphantom{^{***}}[0.0,3.4]^*\hphantom{^{**}}\) & --    & \(\hphantom{^{***}}[0.0,3.3]^{**}\hphantom{^{*}}\) \\[1mm]
    \bottomrule
  \end{tabular}
          \begin{tablenotes}
          \footnotesize
           \item[] \(\!\!\!\!\)\(>3\sigma\) only if:  \quad\,\,
           {\normalsize \(^*\)}cusps differ, \quad\,\,
           {\normalsize \(^{**}\)}\((\tau_u,\tau_d) \simeq (\omega+1,\omega)\), \quad\,\,
           {\normalsize \(^{***}\)}\((\tau_u,\tau_d) \simeq (\omega,\omega)\).
        \end{tablenotes}
  \vskip 1mm
  \caption{Models with \(q^c\sim \mathbf{3}^*\)}
  \label{tab:twob}
  \end{subtable}
        \end{threeparttable}
  \vskip 1mm
  \caption{
  Values and ranges of \(\sqrt{\chi^2}\) for quark fits of 2-moduli 12-parameter (gCP) models, depending on which fields are taken as triplets of \(S_4'\). 
  In~\cref{tab:twoa}, fits below \(9\sigma\) are only possible for both moduli near the same cusp.
  In~\cref{tab:twob}, entries show the minimum and maximum fit \(\sqrt{\chi^2}\) across the 4 different possibilities \((\tau_u,\tau_d) \simeq (\omega,\omega),\,(\omega,\omega+1),\,(\omega+1,\omega),\,(\omega+1,\omega+1)\).
    }
\end{table}
%%%%%%%%%%%%%%%%%%%%%%%%%%%%%%

Our results are summarised in~\cref{tab:twoa,tab:twob}. It is interesting to note that some of the models can fit all the quark data at less than \(3\sigma\), including the CPV phase \(\delta_\text{CP}\), with the moduli being the only source of CP violation (all other parameters are real). Results vary depending on which cusps (left vs.~right) \(\tau_u\) and \(\tau_d\) approach. There are four options for each of the considered cases (\(Q\sim \mathbf{3}^*\) vs.~\(q^c \sim \mathbf{3}^*\)). Overall, one notices that fits are typically better whenever both moduli are in the vicinity of the same cusp.

%%%%%%%%%%%%%%%%%%%%%%%%%%%%%%%%%%%%%%%%%%%%%%
\section{A closer look at natural hierarchies}
\label{sec:natural}
%%%%%%%%%%%%%%%%%%%%%%%%%%%%%%%%%%%%%%%%%%%%%%

In the previous section we checked whether the proposed quark models could fit the known quark data for values of \(\tau\) close to the cusps. 
Even if a good fit is possible, it may be that the proximity of \(\tau\) to a point of residual symmetry is not the main driver behind quark mass hierarchies, since superpotential parameters were not constrained.

\vskip 2mm

In what follows, we analyse a particular model in more detail, looking into potential sources of fine-tuning, namely i) hierarchies between superpotential parameters, which depend heavily on normalisation choices, and ii) cancellations between superpotential parameters. The latter can be analysed e.g.~via a Barbieri-Giudice (BG) measure of fine-tuning~\cite{Barbieri:1987fn}.
We perform this analysis for the model with both \(M_u\) and \(M_d\) taking the form \(M_1\), in the vicinity of the left cusp \(\omega\). In what follows, we still denote the superpotential constants as \(\alpha_i\) in the up sector, while for the down sector we use the notation \(\beta_i\) instead.
Our results are summarised in~\cref{tab:benchmark}.

\clearpage

We consider benchmarks for the following five cases:
\begin{itemize}
\item \textbf{gCP (masses)}: gCP is imposed, but only mass ratios are considered in the fit,
\item \textbf{gCP (all)}:  gCP is imposed (all observables are considered in the fit),
\item \textbf{pheno phase}: a phenomenological phase is added to the up-quark sector,
\item \textbf{no gCP}: a fit of the model in the absence of gCP (two new phases), and
\item \textbf{two moduli}: a fit of the model with gCP imposed, in the presence of an extra modulus (\(\tau_{u,d} = \tau_{1,2}\)).
\end{itemize}
As anticipated from the previous discussion, in the presence of gCP and with a single modulus, one is able to fit quark mass ratios but not all quark data satisfactorily. In particular, the observed strength of CP violation cannot be accommodated. This can be remedied by introducing a single (phenomenological) phase, e.g.,~for \(\alpha_4\) in the up sector, as shown in the third data column of~\cref{tab:benchmark}. It follows that a fit is also possible in the absence of gCP, with independent phases for \(\alpha_4\) and \(\beta_4\), in the up and down sectors respectively (see the fourth data column). Finally, in the presence of gCP, a fit of all quark data --- including CP violation --- is possible with two moduli, one for each sector (recall the results of~\cref{sec:2tau}).

%
%%%%%%%%%%%%%%%%%%%%%%%%%%%%%%
\begin{table}[p]
  \hspace{-1cm}
  \begin{tabular}{lccccc}
    \toprule
     & \textbf{gCP (masses)} & \textbf{gCP (all)} & \textbf{pheno phase} & \textbf{no gCP} & \textbf{two moduli}\\
    \midrule
    \(\re \tau_1\)   & \(-0.4772\) & \(-0.4823\) & \(-0.4992\) & \(-0.4978\) & \(-0.4969\) \\
    \(\im \tau_1\)   & 0.8861 & 0.8784 & 0.8852 & 0.8850 & 0.8692 \\
    \(\re \tau_2\)       & -- & -- & -- & -- & \(-0.4939\) \\
    \(\im \tau_2\)       & -- & -- & -- & -- & 0.8856 \\
    \(|\epsilon_1|\) & 0.0486 & 0.0348 & 0.0306 & 0.0306 & 0.0072 \\
    \(|\epsilon_2|\) & -- & -- & -- & -- & 0.0328 \\
    \midrule
    \(\dfrac{\alpha_2\, \| Y^{(4)}_\mathbf{3'}\|}{\alpha_1\, \| Y^{(1)}_\mathbf{\hat{3}}\|}\) 
                       & 2.725 & 0.009 & 70.93 & 35.79 
                       & 0.329 \\
    \(\dfrac{\alpha_3\, \| Y^{(7)}_{\mathbf{3'},1}\|}{\alpha_1\, \| Y^{(1)}_\mathbf{\hat{3}}\|}\)
                       & 2.128 & 2.975 & 214.0 & 54.52 
                       & 4.341 \\
    \(\dfrac{\alpha_4\, \| Y^{(7)}_{\mathbf{3'},2}\|}{\alpha_1\, \| Y^{(1)}_\mathbf{\hat{3}}\|}\)
                       & 41.86 & 3.979 & \(211.0\, e^{0.890\, i}\) & \(141.5\,e^{-0.356\,i}\) 
                       & 2.555 \\
    \(\dfrac{\beta_2 \, \| Y^{(4)}_\mathbf{3'}\|}{\beta_1 \, \| Y^{(1)}_\mathbf{\hat{3}}\|}\) 
                       & 3.001 & 5.700 & 7.689 & 5.360 
                       & 2.806 \\
    \(\dfrac{\beta_3 \, \| Y^{(7)}_{\mathbf{3'},1}\|}{\beta_1 \, \| Y^{(1)}_\mathbf{\hat{3}}\|}\)
                       & 6.261 & 0.729 & 2.261 & 1.098 
                       & 0.578 \\
    \(\dfrac{\beta_4 \, \| Y^{(7)}_{\mathbf{3'},2}\|}{\beta_1 \, \| Y^{(1)}_\mathbf{\hat{3}}\|}\)
                       & 1.174 & 1.433 & 1.043 & \(1.337\, e^{0.676\,i}\) 
                       & 0.258 \\
    \midrule
    \((m_u/m_c) \,/\, 10^{-3}\) & 2.042 & 1.897 & 2.040 & 2.042 & 2.041 \\
    \((m_c/m_t) \,/\, 10^{-3}\) & 2.678 & 1.824 & 2.678 & 2.678 & 2.678 \\
    \((m_d/m_s) \,/\, 10^{-2}\) & 5.053 & 5.059 & 5.053 & 5.053 & 5.052 \\
    \((m_s/m_b) \,/\, 10^{-2}\) & 1.370 & 1.390 & 1.370 & 1.370 & 1.370 \\
    \(\theta_{12} \,(\degree)\) & 15.42 & 13.05 & 13.03 & 13.03 & 13.03 \\
    \(\theta_{23} \,(\degree)\) & 10.00 & \(4.08\times 10^{-5}\)& 2.055 & 2.054 & 2.054 \\
    \(\theta_{13} \,(\degree)\) & 1.226 & 0.208 & 0.180 & 0.180 & 0.180 \\
    \(\delta_\text{CP} \,(\degree)\) & 0.0026 & 69.24 & 69.21 & 69.21 & 69.21 \\
    \(J_\text{CP} \,/\, 10^{-5}\)    & 0.0042 & \(5.33\times 10^{-5}\) & 2.314 & 2.313 & 2.313 \\
    \midrule
    \((m_c/m_t) \, / \, |\epsilon|\)   & 0.055 & 0.052 & 0.088 & 0.088 & 0.371 \\
    \((m_u/m_t) \, / \, |\epsilon|^2\) & 0.002 & 0.003 & 0.006 & 0.006 & 0.105 \\
    \((m_s/m_b) \, / \, |\epsilon|\)   & 0.282 & 0.400 & 0.448 & 0.448 & 0.418 \\
    \((m_d/m_b) \, / \, |\epsilon|^2\) & 0.293 & 0.582 & 0.739 & 0.739 & 0.643 \\
    \midrule
    max BG \({m_c/m_t}\) & 0.998 & 1.144 & 0.988 & 1.022 & 4.084 \\
    max BG \({m_u/m_t}\) & 2.004 & 1.000 & 0.972 & 1.862 & 2.173 \\
    max BG \({m_s/m_b}\) & 0.841 & 1.229 & 1.089 & 0.759 & 1.167 \\
    max BG \({m_d/m_b}\) & 1.018 & 2.096 & 1.098 & 0.959 & 1.211 \\
    \midrule
    \(N\sigma\) (masses) &  0.0 & 3.4 & 0.0 & 0.0 & 0.0 \\
    \(N\sigma\) (angles) & 51.8 & 5.5 & 0.0 & 0.0 & 0.0 \\
    \(N\sigma\) (\(\delta_\text{CP}\))    & 11.2 & 0.0 & 0.0 & 0.0 & 0.0 \\
    \(N\sigma\) (total)  & 52.9 & 6.4 & 0.0 & 0.0 & 0.0 \\
    \bottomrule
  \end{tabular}
  \caption{Fit benchmarks (see text) for the model with both \(M_{u,d} \sim M_1\). Here, \(\tau_1 = \tau\) and \(\epsilon_1 = \tau\), while \(\tau_{1,2} = \tau_{u,d}\) and \(\epsilon_{1,2} = \epsilon_{u,d}\) in the two-moduli case. All moduli are in the vicinity of the left cusp.
  }
  \label{tab:benchmark}
\end{table}
%%%%%%%%%%%%%%%%%%%%%%%%%%%%%%
%

\subsection{Hierarchical parameters}
\label{sec:hier}
If the proximity to the cusp, i.e.~the smallness of \(|\epsilon|\), is to explain quark mass hierarchies, one expects superpotential parameters to be of the same order, within each sector.
However, as discussed in~\cref{sec:norms}, these parameters may absorb the different choices of modular form normalisations. Therefore, we report the (ratios between) products of superpotential parameters by the Euclidean norm of the corresponding modular form, \(\alpha_i \|Y_i\|\). These products are what (partly) determines the magnitude of the columns of mass matrices.
Requiring superpotential parameters to be of the same order then means that the aforementioned ratios should be \(\mathcal{O}(1)\).
If one chooses to normalise the forms using the Euclidean norm at the fit value of the modulus (moduli), then all \(\|Y_i\|=1\) in~\cref{tab:benchmark}.

The magnitude of the columns of mass matrices is also affected by the canonical normalisation of fields. As mentioned in~\cref{sec:results}, bringing the kinetic terms to a canonical form results in a rescaling of each contribution to the mass matrix by a factor of \(\sqrt{(2 \im \tau)^{k_{Y_i}}}\).%
\footnote{%
In the two-moduli case, we take these factors to be \(\sqrt{(2 \im \tau_1)^{k_{Y_i}}(2 \im \tau_2)^{k_{Y_i}}}\), inspired by the diagonal (\(\tau_3=0\)) scenario in~\cite{Ding:2021iqp}. Note that, in the absence of a complete model, this is a purely heuristic choice.}
Recall that the reported \(\alpha_i\) and \(\beta_i\) are defined via~\mbox{\cref{eq:matricesform}}, prior to taking into account this effect. These additional factors are tied to the specificities of the model and may play a role in naturally enhancing or suppressing mass hierarchies.

\vskip 2mm

As noted in the previous sections, it is quite restrictive to use a single modulus and, correspondingly, a single value of \(|\epsilon|\), common to both quark sectors. Some extra hierarchy in the parameters may be necessary to accommodate all mass ratios, as evidenced by the ``gCP (masses)'' benchmark (first data column of~\cref{tab:benchmark}). Here, one can check that \(\ta_4 \gg \ta_2, \ta_3\), resulting in an extra suppression of mass ratios in the up sector, now controlled by powers of \(|\epsilon / \ta_4|\), as anticipated in~\cref{sec:matrices}.
By adding mixing and CPV constraints to the fit, the limit of interest becomes less transparent, as one may be driven to regions of parameter space with small \(\ta_2\) (``gCP (masses)'' benchmark, second column) or small \(\ta_1\) (``pheno phase'' and ``no gCP'' benchmarks, third and fourth columns).

\subsection{The role of \texorpdfstring{\(|\epsilon|\)}{|ϵ|} and possible cancellations}
\label{sec:canc}

The above single-modulus benchmarks feature values of \(|\epsilon| \sim 0.03-0.05\), whereas for the two-moduli case one can fit the data with \(|\epsilon_1| \sim 0.01\) and \(\epsilon_2 \sim 0.03\).
A simple way to inspect how hierarchies are controlled by the appropriate powers of \(|\epsilon|\) is to look into the ratios 
\begin{equation} \label{eq:eratios}
    (m_c/m_t) \, / \, |\epsilon|\,,\quad
    (m_u/m_t) \, / \, |\epsilon|^2\,,\quad
    (m_s/m_b) \, / \, |\epsilon|\,,\quad
    (m_d/m_b) \, / \, |\epsilon|^2\,,
\end{equation}
which we report in~\cref{tab:benchmark} for each benchmark. These ratios are non-linear functions of the parameters, encoding also the effects of canonical field normalisation, cf.~\cref{eq:M1ratios1,eq:M1ratios2,eq:M1ratios3}.
One expects these ratios to be \(\mathcal{O}(1)\) if the proximity to the cusp is to single-handedly explain the quark hierarchies. The fact that these values are relatively small in the up sector for the single-modulus benchmarks shows that either hierarchies or cancellations of superpotential constants, together with the effect of canonical field rescalings, play an important role in driving the up-quark mass hierarchies.
An exception is the two-moduli case, where the values of these ratios can be milder, thanks to the freedom in varying separately \(\tau_u\) and \(\tau_d\).

One way to gauge how reliant a model is on parameter-driven cancellations is to compute the Barbieri-Giudice (BG) measure of fine-tuning \cite{Barbieri:1987fn}, which allows one to identify regions of parameter space where small changes lead to large deviations in model predictions.
We employ the definition
\begin{align}
\text{max BG } \frac{m_i}{m_j} \, \equiv\,
\max_{\substack{p \,=\, \alpha,\beta\\ k \,=\, 2,3,4}} \,\left|\frac{\partial \ln m_i/m_j}{\partial \ln p_k/p_1}\right|\,,
\end{align}
singling out the largest effect on mass ratios across superpotential parameters. 
Note that this measure is not suitable for angular variables. BG values are reported in~\cref{tab:benchmark} for the benchmarks above. Overall, one finds acceptable values with the exception perhaps of the two-moduli benchmark, featuring an apparently tuned ratio \(m_c / m_t\).
We have checked that this effect is driven by quark mixing and CP violation, to the extent that a fit of quark mass ratios alone is possible in this scenario, for similar values of \(|\epsilon_u|\) and \(|\epsilon_d|\), with all BG \(\sim 1\), with \(\mathcal{O}(0.1-1)\) ratios of \(\alpha_i \|Y_i\|\), and with the quantities in~\cref{eq:eratios} within the interval \([0.2,0.7]\).

%%%%%%%%%%%%%%%%%%%%%%%%%%%%%%%%%%%%%%%%%%%%%%
\section{Summary and conclusions}
\label{sec:summary}
%%%%%%%%%%%%%%%%%%%%%%%%%%%%%%%%%%%%%%%%%%%%%%

Obtaining an economical and fine-tuning-free description of the quark sector --- i.e.~of quark mass
hierarchies, mixing and CP violation --- within the modular flavour approach still remains a serious challenge.
In this work, we have spelled out the challenges for building viable and minimal \(S_4'\) modular-invariant quark models where the proximity to the point of residual \(\mathbb{Z}_3^{ST}\) symmetry plays a role in determining mass hierarchies, via powers of a small parameter \(|\epsilon|\)~\cite{Novichkov:2021evw}.

We argue that, in a bottom-up approach, the absolute normalisations of the modular forms are arbitrary and should not determine hierarchies. These can instead follow from the relative magnitudes of modular multiplet components.
We thus identify four minimal mass matrix patterns, \(M_{1,2}\), \(M_3\), \(M_4\) and \(M_{5,6}\), where the quark mass hierarchies may stem from the proximity of \(\tau\) to the cusp (the smallness of \(|\epsilon|\)). These involve at most four superpotential parameters and do not lead to massless fermions.
Approximate analytical expressions have been derived for the quark masses and mass ratios, for each of these structures. These results are directly applicable to other fermions, e.g.~the charged lepton sector.

Two main hurdles to overcome in this class of models are i) the different hierarchies observed in the up and down sectors, which call for different values of \(|\epsilon|\), and ii) the suppression of CP violation whenever \(\tau\) is the only source of CP symmetry breaking, already alluded to in Ref.~\cite{Petcov:2022fjf}.  
The former indicates that, to explain quark mass ratios in the single-modulus case, one may need to tolerate some hierarchy among superpotential couplings.
As for the latter, we find that the observed strength of quark CP violation cannot be adequately fitted in the minimal 10-parameter scenarios, featuring a gCP symmetry.
Instead, we are able to fit quark data with 11 parameters in a phenomenological approach, by explicitly adding a complex phase in one of the sectors. The consistent lifting of the gCP symmetry leads to models with 12 parameters --- one less than previous constructions in the literature --- which can also fit quark data.
Note that having as many parameters as observables does not automatically guarantee a viable fit (see e.g.~\cref{tab:constrainedb}).
Curiously, a 12-parameter fit including the correct amount of CP violation can be achieved near the cusp and in the presence of gCP, provided different moduli are responsible for each of the quark sectors.

In summary, these results illustrate how the demand for explanatory power and for the absence of different kinds of tuning may restrict models of flavour based on modular invariance. It is hoped that these requirements select only a few viable models that provide a clear understanding of the puzzling flavour structures of fundamental fermions.

%%%%%%%%%%%%%%%%%%%%%%%%%%%%%%%%%%%%%%%%%%%%%%
\section*{Acknowledgements}
%%%%%%%%%%%%%%%%%%%%%%%%%%%%%%%%%%%%%%%%%%%%%%

J.T.P.~and M.L.~would like to thank A.~Azatov and acknowledge the MIUR grant PRIN 2017L5W2PT for partial financial support during their visit to SISSA.
This work was partially supported by Fundação para a Ciência e a Tecnologia (FCT, Portugal) through the projects~CERN/FIS-PAR/0002/2021,~CERN/FIS-PAR/0019/2021, and CFTP-FCT Unit~UIDB/00777/2020 and~UIDP/00777/2020, which are partially funded through POCTI (FEDER), COMPETE, QREN and EU.
IdMV and J.T.P.~acknowledge support by FCT fellowships in the framework of the project UIDP/00777/2020.
M.L.~acknowledges support from FCT through the grant No.~PD/BD/150488/2019, in the framework of the Doctoral Programme IDPASC-PT.
The work of S.T.P.~was supported in part by the European Union's Horizon 2020 research and innovation programme under the Marie Sk\l{}odowska-Curie grant agreement No.~860881-HIDDeN, by the Italian INFN program on Theoretical Astroparticle Physics and by the World Premier International Research Center Initiative (WPI Initiative, MEXT), Japan.

\appendix
%%%%%%%%%%%%%%%%%%%%%%%
\section{Modular \texorpdfstring{$S_4'$}{S4'}}
\label{app:modS4p}
%%%%%%%%%%%%%%%%%%%%%%%

%%%%%%%%%%%%%%%%%%%%%%%
\subsection{Group theory}
\label{app:group}
%%%%%%%%%%%%%%%%%%%%%%%
The homogeneous finite modular group \(S_4' \equiv SL(2,\mathbb{Z}_4)\), with group ID \texttt{[48, 30]} in the computer algebra system GAP~\cite{GAP4,SmallGroups}, 
is a group of 48 elements defined by the three generators \(S\), \(T\) and \(R\) satisfying:
\begin{align}
S^2 = R\,,\quad
T^4 = (ST)^3 = R^2 = \id\,, \quad
TR = RT
\,.
\end{align}
It admits 10 irreducible representations, denoted by
\begin{align}
    \mathbf{1}\,,\,
    \mathbf{\hat{1}}\,,\,
    \mathbf{1'}\,,\,
    \mathbf{\hat{1}'}\,,\,
    \mathbf{2}\,,\,
    \mathbf{\hat{2}}\,,\,
    \mathbf{3}\,,\,
    \mathbf{\hat{3}}\,,\,
    \mathbf{3'}\,,\,
    \mathbf{\hat{3}'}\,,
\end{align}
where irreps without a hat have a direct correspondence with \(S_4\) irreps. 
The working basis for the representation matrices of the group generators coincides with the one used in Ref.~\cite{Novichkov:2020eep}, where also Clebsch-Gordan coefficients can be found. 
The chosen basis is symmetric and thus convenient for the study of modular symmetry extended by a gCP symmetry~\cite{Novichkov:2019sqv,Novichkov:2020eep}.

%%%%%%%%%%%%%%%%%%%%%%%
\subsection{Modular forms}
\label{app:mod_forms}
%%%%%%%%%%%%%%%%%%%%%%%

Modular multiplets for the homogeneous finite modular group \(S_4'\) can be written in terms of the two functions \(\theta(\tau)\) and \(\varepsilon(\tau)\) defined in~\cref{eq:theta_eps_qexp},
and can be found in section 3 and appendix D of Ref.~\cite{Novichkov:2020eep}, for weights up to \(k=10\). We make use of the modular multiplets \(Y_{\mathbf{r}}^{(k)}(\tau)\) given therein, which we reproduce here for convenience, up to weight \(k=8\). Note that the dimensionality of the linear spaces of \(S_4'\) modular forms of weight \(k\) is given by \(2k+1\), in agreement with these results.

Of particular relevance to our study is an \(ST\)-diagonal basis, where the power structure in \(\epsilon \sim \tau-\omega\) becomes apparent. Such a basis can be defined by \(\rho_{\mathbf{2}^*}(ST) = \diag(\omega,\omega^2)\) and \(\rho_{\mathbf{3}^*}(ST) = \diag(1,\omega,\omega^2)\) for all doublets \(\mathbf{2}^*\) and triplets \(\mathbf{3}^*\) of \(S_4'\). For modular forms in these representations, the relative suppressions of modular multiplet elements are physically significant, as argued in~\cref{sec:norms}. Therefore, we additionally present the profile of modular doublets and triplets in the \(ST\)-diagonal basis, in the vicinity of the cusp \(\omega\), ignoring multiplet normalisation, i.e.~the \(\mathcal{O}(\epsilon^0)\) term is scaled to 1, and keeping only the leading term in each entry. This approximate proportionality is denoted by the symbol ``\(\aprop\)''. Some modular forms vanish at \(\tau = \omega\), e.g.~\(Y_{\mathbf{2}}^{(6)}\) for which the overall factor of \(\epsilon\) is shown explicitly.

Note finally that whenever there is more than one multiplet for the same weight \(k\) and irrep \(\mathbf{r}\), one is also free to choose the basis of the corresponding subspace --- in a bottom-up approach there is no known rule to select which are the ``correct'' linear combinations of the multiplets one should consider. Moreover, one of these combinations may vanish at a certain \(\tau\) (e.g.~the cusp).
While the triplets of interest, as defined in~\cite{Novichkov:2020eep}, do not vanish at \(\tau = \omega\), one may find a linear combination of these forms (with the same \(k\) and \(\mathbf{r}\)) that does. In this specific basis, normalising the \(\epsilon\)-suppressed form may then suggest a different \(\epsilon\) power structure for the triplet. Such a basis choice may be a source of fine-tuning, and we do not consider it further in this work.

\subsubsection{Weight 1}
\begin{align*}
  Y_{\mathbf{\hat{3}}}^{(1)}(\tau) &=
  \begin{pmatrix}
    \sqrt{2} \, \varepsilon \, \theta \\[1mm]
    \varepsilon^2 \\[1mm]
    -\theta^2
  \end{pmatrix}
  \ato
    \begin{pmatrix}
    \frac{1}{\sqrt{3}}\,\epsilon \\[1mm]
    1 \\[1mm]
    -\frac{1}{6}\,\epsilon^2 
  \end{pmatrix}
  \,.
\end{align*}

\subsubsection{Weight 2}
\begin{align*}
  Y_{\mathbf{2}}^{(2)}(\tau) &=
  \begin{pmatrix}
    \frac{1}{\sqrt{2}}\left(\theta^4 + \varepsilon^4\right) \\[1mm]
    -\sqrt{6}\, \varepsilon^2\, \theta^2
  \end{pmatrix}
  \ato
    \begin{pmatrix}
    -\frac{2}{\sqrt{3}}\,\epsilon \\[1mm]
    1
  \end{pmatrix}
  \,,
  \\[2mm]
    Y_{\mathbf{3'}}^{(2)}(\tau) &=
  \begin{pmatrix}
    \frac{1}{\sqrt{2}}\left(\theta^4 - \varepsilon^4\right) \\[1mm]
    -2 \,\varepsilon \, \theta^3\\[1mm]
    -2 \,\varepsilon^3 \, \theta
  \end{pmatrix}
  \ato
    \begin{pmatrix}
    -\frac{1}{2}\,\epsilon^2 \\[1mm]
    \frac{1}{\sqrt{3}}\,\epsilon\\[1mm]
    1
  \end{pmatrix}
  \,.
\end{align*}

\subsubsection{Weight 3}
\begin{align*}
  Y_{\mathbf{\hat{1}'}}^{(3)}(\tau) &= \sqrt{3} \left(
  \varepsilon \,\theta^5-\varepsilon^5\, \theta
  \right)\,,
  \\[2mm]
  Y_{\mathbf{\hat{3}}}^{(3)}(\tau) &=
  \begin{pmatrix}
 \varepsilon ^5\, \theta +\varepsilon \, \theta ^5\\[1mm]
 \frac{1}{2\sqrt{2}}\left(5 \,\varepsilon ^2 \, \theta ^4-\varepsilon ^6 \right)\\[1mm]
 \frac{1}{2\sqrt{2}}\left(\theta ^6-5\, \varepsilon ^4 \,\theta ^2\right)
  \end{pmatrix}
  \ato
  \begin{pmatrix}
  1\\[1mm]
 \frac{5}{6}\,\epsilon^2 \\[1mm]
 -\frac{1}{\sqrt{3}}\,\epsilon
  \end{pmatrix}
  \,,
  \\[2mm]
    Y_{\mathbf{\hat{3}'}}^{(3)}(\tau) &= \frac{1}{2}
  \begin{pmatrix}
 -4 \sqrt{2}\, \varepsilon ^3 \,\theta ^3 \\[1mm]
 \theta ^6 + 3 \,\varepsilon ^4\, \theta ^2 \\[1mm]
 - 3\, \varepsilon ^2\, \theta ^4 -\varepsilon ^6
  \end{pmatrix}
  \ato
  \begin{pmatrix}
  1\\[1mm]
 -\frac{1}{2}\,\epsilon^2 \\[1mm]
 \sqrt{3}\, \epsilon
  \end{pmatrix}
  \,.
\end{align*}

\subsubsection{Weight 4}
\begin{align*}
  Y_{\mathbf{1}}^{(4)}(\tau) &=
  \frac{1}{2 \sqrt{3}} \left(
  \theta^8 + 14\, \varepsilon^4\, \theta^4 + \varepsilon^8
  \right)\,,
  \\[2mm]
  Y_{\mathbf{2}}^{(4)}(\tau) &=
  \begin{pmatrix}
  \frac{1}{4} \left(\theta^8 - 10 \,\varepsilon^4\, \theta^4 + \varepsilon^8\right) \\[1mm]
  \sqrt{3}\left(\varepsilon ^2\,\theta^6 + \varepsilon ^6\, \theta^2\right)
  \end{pmatrix}
  \ato
  \begin{pmatrix}
  1 \\[1mm] -\frac{4}{3}\,\epsilon^2
  \end{pmatrix}
  \,,
  \\[2mm]
  Y_{\mathbf{3}}^{(4)}(\tau) &=
  \frac{3}{2\sqrt{2}}
  \begin{pmatrix}
 \sqrt{2}\left(\varepsilon ^2\, \theta^6 -\varepsilon^6 \, \theta ^2\right)\\[1mm]
 \varepsilon ^3 \,\theta ^5 - \varepsilon ^7 \,\theta \\[1mm]
 - \varepsilon  \,\theta ^7 + \varepsilon ^5 \,\theta ^3
  \end{pmatrix}
  \ato
  \begin{pmatrix}
  \frac{1}{\sqrt{3}} \, \epsilon \\[1mm] 1\\[1mm]- \frac{1}{6}\,\epsilon^2
  \end{pmatrix}
  \,,
  \\[2mm]
  Y_{\mathbf{3'}}^{(4)}(\tau) &=
  \begin{pmatrix}
  \frac{1}{4}\left(\theta ^8-\varepsilon ^8\right)\\[1mm]
  \frac{1}{2\sqrt{2}}\left(\varepsilon \, \theta ^7 + 7 \,\varepsilon ^5\, \theta ^3\right) \\[1mm]
  \frac{1}{2\sqrt{2}}\left(7 \,\varepsilon ^3\, \theta ^5 + \varepsilon ^7 \,\theta\right)
  \end{pmatrix}
    \ato
  \begin{pmatrix}
  - \sqrt{3} \, \epsilon \\[1mm] 1\\[1mm] \frac{7}{6}\,\epsilon^2
  \end{pmatrix}
  \,.
\end{align*}

\subsubsection{Weight 5}
\begin{align*}
  Y_{\mathbf{\hat{2}}}^{(5)}(\tau) &=
  \begin{pmatrix}
  \frac{3}{2}\left( \varepsilon ^3 \,\theta ^7-\varepsilon ^7 \,\theta ^3 \right)\\[1mm]
  \frac{\sqrt{3}}{4}\left( \varepsilon  \,\theta ^9-\varepsilon ^9 \,\theta   \right)
  \end{pmatrix}
  \ato
  \begin{pmatrix}
  \frac{2}{\sqrt{3}}\, \epsilon \\[1mm] 1
  \end{pmatrix}
  \,,
  \\[2mm]
  Y_{\mathbf{\hat{3}},1}^{(5)}(\tau) &=
  \begin{pmatrix}
  \frac{6\sqrt{2}}{\sqrt{5}} \,
\varepsilon ^5\, \theta ^5
\\[1mm]
\:\:\:\frac{3}{8\sqrt{5}} \left(
5\, \varepsilon ^2\, \theta ^8
+10\, \varepsilon ^6\, \theta ^4
+\varepsilon ^{10}
\right) \\[1mm]
-\frac{3}{8\sqrt{5}} \left(
\theta ^{10}
+10\, \varepsilon ^4\, \theta ^6
+5\, \varepsilon ^8\, \theta ^2
\right)
  \end{pmatrix}
  \ato
  \begin{pmatrix}
    - \frac{5}{2}\,\epsilon^2 \\[1mm] -\frac{5}{\sqrt{3}}\, \epsilon \\[1mm] 1
  \end{pmatrix}\,,
  \\[2mm]
  Y_{\mathbf{\hat{3}},2}^{(5)}(\tau) &=
  \begin{pmatrix}
  \frac{3}{4}\left( \varepsilon  \, \theta ^9 -2 \,\varepsilon ^5 \, \theta ^5 + \varepsilon ^9 \, \theta \right)\\[1mm]
  \frac{3}{\sqrt{2}}\left(-\varepsilon ^2 \, \theta ^8 + \varepsilon ^6 \, \theta ^4\right)\\[1mm]
  \frac{3}{\sqrt{2}}\left(-\varepsilon ^4 \, \theta ^6 + \varepsilon ^8 \, \theta ^2\right)
  \end{pmatrix}
    \ato
  \begin{pmatrix}
    - \frac{1}{2}\,\epsilon^2 \\[1mm] \frac{1}{\sqrt{3}}\, \epsilon \\[1mm] 1
  \end{pmatrix}
  \,,
  \\[2mm]
  Y_{\mathbf{\hat{3}'}}^{(5)}(\tau) &=
  \begin{pmatrix}
  2\left(\varepsilon ^3\, \theta ^7 +  \varepsilon ^7 \,\theta ^3\right)\\[1mm]
  \frac{1}{4\sqrt{2}}\left(\theta ^{10} -14 \,\varepsilon ^4\, \theta ^6  -3 \,\varepsilon ^8\, \theta ^2 \right) \\[1mm]
  \frac{1}{4\sqrt{2}}\left( 3\, \varepsilon ^2\, \theta ^8 + 14\, \varepsilon ^6\, \theta ^4 -\varepsilon ^{10 }\right)
  \end{pmatrix}
      \ato
  \begin{pmatrix}
    \frac{3}{2}\,\epsilon^2 \\[1mm] -\frac{1}{\sqrt{3}}\, \epsilon \\[1mm] 1
  \end{pmatrix}
  \,.
\end{align*}

\subsubsection{Weight 6}
\begin{align*}
  Y_{\mathbf{1}}^{(6)}(\tau) &=
  \frac{1}{4\sqrt{6}} \left(
\theta ^{12}
-33 \,\varepsilon ^4 \,\theta ^8
-33 \,\varepsilon ^8 \,\theta ^4
+\varepsilon ^{12}
\right)
  \,,\\[2mm]
  Y_{\mathbf{1'}}^{(6)}(\tau) &=
  \frac{3}{2}\sqrt{\frac{3}{2}} \left(
\varepsilon ^2\, \theta ^{10}
-2 \,\varepsilon ^6\, \theta ^6
+ \varepsilon ^{10} \,\theta ^2
\right)
  \,,\\[2mm]
  Y_{\mathbf{2}}^{(6)}(\tau) &=
  \begin{pmatrix}
  \frac{1}{8} \left(
\theta ^{12}
+15\, \varepsilon ^4\, \theta ^8
+15\, \varepsilon ^8\, \theta ^4
+\varepsilon ^{12}
\right) \\[1mm]
-\frac{\sqrt{3}}{4} \left(
\varepsilon ^2\, \theta ^{10}
+14\, \varepsilon ^6\, \theta ^6
+\varepsilon ^{10}\, \theta ^2
\right)
  \end{pmatrix}
\ato
\epsilon
\begin{pmatrix}
-\frac{2}{\sqrt{3}} \, \epsilon \\[1mm] 1
\end{pmatrix}
\,,
  \\[2mm]
  Y_{\mathbf{3}}^{(6)}(\tau) &=
  \begin{pmatrix}
\frac{3}{2}
\left(\varepsilon ^2 \,\theta ^{10}-\varepsilon ^{10}\, \theta ^2 \right)
		 \\[1mm]
\frac{3}{4\sqrt{2}} \left(
 5\, \varepsilon ^3\, \theta ^9
 -6\, \varepsilon ^7\, \theta ^5
 +\varepsilon ^{11} \,\theta
		\right) \\[1mm]
\frac{3}{4\sqrt{2}} \left(
 \varepsilon\,  \theta ^{11}
 -6 \,\varepsilon ^5\, \theta ^7
 +5 \,\varepsilon ^9\, \theta ^3
		\right)
  \end{pmatrix}
\ato
\begin{pmatrix}
1 \\[1mm] \frac{5}{6} \, \epsilon^2 \\[1mm] - \frac{1}{\sqrt{3}} \, \epsilon
\end{pmatrix}  
  \,,
  \\[2mm]
    Y_{\mathbf{3'},1}^{(6)}(\tau) &=
  \begin{pmatrix}
  -\frac{3}{8\sqrt{13}}\left(
\theta ^{12}
 -3\, \varepsilon ^4\, \theta ^8
 +3\, \varepsilon ^8\, \theta ^4
 -\varepsilon ^{12}
\right)
\\[1mm]
\frac{3 \sqrt{2}}{\sqrt{13}}\left(
 3 \,\varepsilon ^5\, \theta ^7
 + \varepsilon ^9\, \theta ^3
\right) \\[1mm]
\frac{3 \sqrt{2}}{\sqrt{13}} \left(
 \varepsilon ^3\, \theta ^9
 + 3\, \varepsilon ^7\, \theta ^5
\right)
  \end{pmatrix}
\ato
\begin{pmatrix}
1 \\[1mm] -\frac{5}{2} \, \epsilon^2 \\[1mm] - \sqrt{3} \, \epsilon
\end{pmatrix}  
\,,
  \\[2mm]
  Y_{\mathbf{3'},2}^{(6)}(\tau) &=
  \begin{pmatrix}
    3\left(
     \varepsilon ^4\, \theta ^8
 -\varepsilon ^8\, \theta ^4
     \right)\\[1mm]
  -\frac{3}{4\sqrt{2}}\left(
 \varepsilon\,  \theta ^{11}
 +2\, \varepsilon ^5\, \theta ^7
 -3\, \varepsilon ^9\, \theta ^3
  \right)\\[1mm]
 \:\:\:\, \frac{3}{4\sqrt{2}}\left(
 3\, \varepsilon ^3\, \theta ^9
 -2\, \varepsilon ^7\, \theta ^5
 -\varepsilon ^{11}\, \theta
  \right)
\end{pmatrix}
\ato
\begin{pmatrix}
1 \\[1mm] -\frac{1}{2} \, \epsilon^2 \\[1mm] \sqrt{3} \, \epsilon
\end{pmatrix}  \,.
\end{align*}

\subsubsection{Weight 7}
\begin{align*}
  Y_{\mathbf{\hat{1}'}}^{(7)}(\tau) &=
  \frac{1}{4} \sqrt{\frac{3}{2}} \left(
-\varepsilon ^{13}\, \theta
-13\, \varepsilon ^9\, \theta ^5
+13\, \varepsilon ^5\, \theta ^9
+\varepsilon\,  \theta ^{13}
\right)
  \,,\\[2mm]
  Y_{\mathbf{\hat{2}}}^{(7)}(\tau) &=
  \begin{pmatrix}\frac{3}{2} \left(
\varepsilon ^3\, \theta ^{11}
-\varepsilon ^{11} \,\theta ^3
\right) \\[1mm]
-\frac{\sqrt{3}}{8} \left(
\varepsilon\,  \theta ^{13}
-11\, \varepsilon ^5\, \theta ^9
+11\, \varepsilon ^9\, \theta ^5
-\varepsilon ^{13} \,\theta
\right)
  \end{pmatrix}
  \ato \begin{pmatrix}
1 \\[1mm] \frac{4}{3}\,\epsilon^2
\end{pmatrix}
  \,,
  \\[2mm]
  Y_{\mathbf{\hat{3}},1}^{(7)}(\tau) &=
  \begin{pmatrix}
  \frac{12}{\sqrt{13}} \left(
\varepsilon ^5\, \theta ^9
 +\varepsilon ^9\, \theta ^5
\right) \\[1mm]
\frac{3}{8\sqrt{26}} \left(
\varepsilon ^2\, \theta ^{12}
 +45\, \varepsilon ^6\, \theta ^8
 +19\, \varepsilon ^{10} \,\theta ^4
 -\varepsilon ^{14}
\right) \\[1mm]
\frac{3}{8\sqrt{26}} \left(
 \theta ^{14}
 -19\, \varepsilon ^4\, \theta ^{10}
 -45\, \varepsilon ^8\, \theta ^6
 -\varepsilon ^{12}\, \theta ^2
\right)
\end{pmatrix}
  \ato \begin{pmatrix}
\sqrt{3}\,\epsilon  \\[1mm]
1 \\[1mm] -\frac{5}{6}\,\epsilon^2
\end{pmatrix}
\,,
  \\[2mm]
  Y_{\mathbf{\hat{3}},2}^{(7)}(\tau) &=
  \begin{pmatrix}
  \frac{3}{8} \left(
 \varepsilon\,  \theta ^{13}
 -\varepsilon ^5\, \theta ^9
 -\varepsilon ^9\, \theta ^5
 +\varepsilon ^{13}\, \theta
\right) \\[1mm]
\frac{3}{4\sqrt{2}} \left(
 \varepsilon ^2\, \theta ^{12}
 +6\, \varepsilon ^6\, \theta ^8
 -7\, \varepsilon ^{10} \,\theta ^4
\right) \\[1mm]
\frac{3}{4\sqrt{2}} \left(
 7\, \varepsilon ^4\, \theta ^{10}
 -6\, \varepsilon ^8\, \theta ^6
 -\varepsilon ^{12} \,\theta ^2
\right)
  \end{pmatrix}
    \ato \begin{pmatrix}
-\sqrt{3}\,\epsilon  \\[1mm]
1 \\[1mm] \frac{7}{6}\,\epsilon^2
\end{pmatrix}
  \,,
  \\[2mm]
  Y_{\mathbf{\hat{3}'},1}^{(7)}(\tau) &=
  \begin{pmatrix}
  \frac{3}{4\sqrt{37}}\left(
  7\, \varepsilon ^3 \,\theta ^{11}
 +50\, \varepsilon ^7\, \theta ^7
 +7\, \varepsilon ^{11}\, \theta ^3
\right) \\[1mm]
-\frac{3}{4\sqrt{74}} \left(
\theta ^{14}
 +14\, \varepsilon ^4\, \theta ^{10}
 +49\, \varepsilon ^8\, \theta ^6
\right) \\[1mm]
\:\:\:\frac{3}{4\sqrt{74}} \left(
49\, \varepsilon ^6\, \theta ^8
 +14\, \varepsilon ^{10} \,\theta ^4
 +\varepsilon ^{14}
\right)
\end{pmatrix}
    \ato \begin{pmatrix}
-\frac{7}{\sqrt{3}}\,\epsilon  \\[1mm]
1 \\[1mm] -\frac{49}{6}\,\epsilon^2
\end{pmatrix}
\,,
  \\[2mm]
  Y_{\mathbf{\hat{3}'},2}^{(7)}(\tau) &=
  \begin{pmatrix}
\frac{9}{4}
\left(  \varepsilon ^3\, \theta ^{11}
 -2\, \varepsilon ^7\, \theta ^7
 +\varepsilon ^{11} \,\theta ^3 \right)
\\[1mm]
\:\:\:\frac{9}{4\sqrt{2}} \left(
\varepsilon ^4\, \theta ^{10}
 -2\, \varepsilon ^8\, \theta ^6
 +\varepsilon ^{12}\, \theta ^2
\right) \\[1mm]
-\frac{9}{4\sqrt{2}} \left(
\varepsilon ^2\, \theta ^{12}
 -2\, \varepsilon ^6\, \theta ^8
 +\varepsilon ^{10} \,\theta ^4
\right)
  \end{pmatrix}
      \ato \begin{pmatrix}
\frac{1}{\sqrt{3}}\,\epsilon  \\[1mm]
1 \\[1mm] -\frac{1}{6}\,\epsilon^2
\end{pmatrix}
  \,.
\end{align*}

\subsubsection{Weight 8}
\begin{align*}
  Y_{\mathbf{1}}^{(8)}(\tau) &=
\frac{1}{8\sqrt{6}}
\left(
 \theta^{16}
+28 \,\varepsilon^4 \,\theta^{12}
+198 \,\varepsilon^8 \,\theta^8
+28 \,\varepsilon^{12} \,\theta^4
+\varepsilon^{16}
\right)
  \,,\\[2mm]
  Y_{\mathbf{2},1}^{(8)}(\tau) &=
  \begin{pmatrix}
\frac{9}{16\sqrt{82}}
\left(
 \theta^{16}
-130 \,\varepsilon^8 \,\theta^8
+\varepsilon^{16}
\right)
\\[1mm]
\frac{3}{8}\sqrt{\frac{3}{82}}
\left(
 5 \,\varepsilon^2 \,\theta^{14}
+91 \,\varepsilon^6 \,\theta^{10}
+91 \,\varepsilon^{10} \,\theta^6
+5 \,\varepsilon^{14} \,\theta^2
\right)
  \end{pmatrix}
  \ato
  \begin{pmatrix}
-\frac{34}{\sqrt{3}}\,\epsilon \\[1mm] 1
\end{pmatrix}  
  \,,
  \\[2mm]
  Y_{\mathbf{2},2}^{(8)}(\tau) &=
  \begin{pmatrix}
\frac{9}{4}
\left(
 \varepsilon^4 \,\theta^{12}
-2 \,\varepsilon^8 \,\theta^8
+\varepsilon^{12} \,\theta^4
\right)
\\[1mm]
\frac{3\sqrt{3}}{8}
\left(
 \varepsilon^2 \,\theta^{14}
-\varepsilon^6 \,\theta^{10}
-\varepsilon^{10} \,\theta^6
+\varepsilon^{14} \,\theta^2
\right)
  \end{pmatrix}
  \ato
  \begin{pmatrix}
\frac{2}{\sqrt{3}}\,\epsilon \\[1mm] 1
\end{pmatrix}  \,,
  \\[2mm]
  Y_{\mathbf{3},1}^{(8)}(\tau) &=
  \begin{pmatrix}
9\sqrt{\frac{2}{5}}
\left(
 \varepsilon^6 \,\theta^{10}
-\varepsilon^{10} \,\theta^6
\right)
\\[1mm]
\:\:\:\,\frac{9}{16\sqrt{5}}
\left(
 5 \,\varepsilon^3 \,\theta^{13}
+5 \,\varepsilon^7 \,\theta^9
-9 \,\varepsilon^{11} \,\theta^5
-\varepsilon^{15} \,\theta
\right)
\\[1mm]
-\frac{9}{16\sqrt{5}}
\left(
 \varepsilon \, \theta^{15}
+9 \,\varepsilon^5 \,\theta^{11}
-5 \,\varepsilon^9 \,\theta^7
-5 \,\varepsilon^{13} \,\theta^3
\right)
  \end{pmatrix}  
  \ato
  \begin{pmatrix}
  -\frac{5}{2}\,\epsilon^2\\[1mm]-\frac{5}{\sqrt{3}}\,\epsilon \\[1mm] 1
\end{pmatrix}  
  \,,
  \\[2mm]
  Y_{\mathbf{3},2}^{(8)}(\tau) &=
  \begin{pmatrix}
-\frac{9}{8}
\left(
 \varepsilon^2 \,\theta^{14}
-3 \,\varepsilon^6 \,\theta^{10}
+3 \,\varepsilon^{10} \,\theta^6
-\varepsilon^{14} \,\theta^2
\right)
\\[1mm]
\frac{9}{2\sqrt{2}}
\left(
\varepsilon^3 \,\theta^{13}
-2 \,\varepsilon^7 \,\theta^9
+\varepsilon^{11} \,\theta^5
\right)
\\[1mm]
\frac{9}{2\sqrt{2}}
\left(
 \varepsilon^5 \,\theta^{11}
-2 \,\varepsilon^9 \,\theta^7
+\varepsilon^{13} \,\theta^3
\right)
  \end{pmatrix}
  \ato
  \begin{pmatrix}
  -\frac{1}{2}\,\epsilon^2\\[1mm]\frac{1}{\sqrt{3}}\,\epsilon \\[1mm] 1
\end{pmatrix}  
\,,
  \\[2mm]
  Y_{\mathbf{3'},1}^{(8)}(\tau) &=
  \begin{pmatrix}
\frac{3}{50}
\left(
 \theta^{16}
-\varepsilon^{16}
\right)
\\[1mm]
\frac{3}{200\sqrt{2}}
\left(
 \varepsilon \, \theta^{15}
+273 \,\varepsilon^5 \,\theta^{11}
+715 \,\varepsilon^9 \,\theta^7
+35 \,\varepsilon^{13} \,\theta^3
\right)
\\[1mm]
\frac{3}{200\sqrt{2}}
\left(
 35 \,\varepsilon^3 \,\theta^{13}
+715 \,\varepsilon^7 \,\theta^9
+273 \,\varepsilon^{11} \,\theta^5
+\varepsilon^{15} \,\theta
\right)
  \end{pmatrix}
  \ato
  \begin{pmatrix}
-\frac{75}{14}\,\epsilon^2  \\[1mm] \frac{41}{7 \sqrt{3}}\,\epsilon \\[1mm] 1  
\end{pmatrix}  \,,
  \\[2mm]
  Y_{\mathbf{3'},2}^{(8)}(\tau) &=
  \begin{pmatrix}
3
\left(
 \varepsilon^4 \,\theta^{12}
-\varepsilon^{12} \,\theta^4
\right)
\\[1mm]
\frac{3}{8\sqrt{2}}
\left(
 \varepsilon \, \theta^{15}
-15 \,\varepsilon^5 \,\theta^{11}
+11 \,\varepsilon^9 \,\theta^7
+3 \,\varepsilon^{13} \,\theta^3
\right)
\\[1mm]
\frac{3}{8\sqrt{2}}
\left(
 3 \,\varepsilon^3 \,\theta^{13}
+11 \,\varepsilon^7 \,\theta^9
-15 \,\varepsilon^{11} \,\theta^5
+\varepsilon^{15} \,\theta
\right)
  \end{pmatrix}
  \ato
  \begin{pmatrix}
\frac{3}{2}\,\epsilon^2  \\[1mm]- \frac{1}{\sqrt{3}}\,\epsilon \\[1mm] 1  
\end{pmatrix}  \,.
\end{align*}

%%%%%%%%%%%%%%%%%%%%%%%
\bibliographystyle{JHEPwithnote}
\bibliography{bibliography}
%%%%%%%%%%%%%%%%%%%%%%%

\end{document}